\documentclass[journal]{IEEEtran}
\usepackage[table]{xcolor}
\usepackage[hidelinks]{hyperref}
\usepackage{cite}

\ifCLASSINFOpdf
  \usepackage[pdftex]{graphicx}
\else
  \usepackage[dvips]{graphicx}
  \DeclareGraphicsExtensions{.eps}
\fi

\usepackage{gnuplot-lua-tikz}
\usepackage{xcolor}
\definecolor{gt@red}{RGB}{216,30,5}
\definecolor{gt@dkred}{RGB}{128,0,0}
\definecolor{gt@gray}{RGB}{110,124,146}
\definecolor{gt@dkgray}{RGB}{98,110,128}
\definecolor{gt@yellow}{RGB}{255,128,0}
\definecolor{gt@blue}{RGB}{0,68,153}

\usepackage{tikz}
\usepackage{xifthen}
\usepackage{adjustbox}
\usepackage{xspace}
\usepackage{csquotes}
\usepackage[nolist]{acronym}

\usepackage{booktabs}
\usepackage{listings}
\usepackage{url}

\newcommand\lt[1]{{\lstinline+#1+}}
\renewcommand\t[1]{{\lstinline+#1+}}

\definecolor{dkgreen}{rgb}{0,0.5,0}
\definecolor{dkred}{rgb}{0.5,0,0}
\definecolor{gray}{rgb}{0.5,0.5,0.5}
\newcommand{\tool}[1]{\textsc{#1}}
\newcommand{\thename}{\textsc{LLAnalyzer}\xspace}
\newcommand{\dbitool}{\textsc{SignalSeer}\xspace}

\newcommand{\numexecutions}{{1778}\xspace}
\newcommand{\numunbalancedcorrect}{{1658}\xspace}
\newcommand{\numunbalancederrors}{{120}\xspace}

\newcommand\multiII[1]{\multicolumn{2}{c}{#1}}

\begin{document}

\begin{acronym}
\acro{QA}{Quality Assurance}
\acro{SUT}{System Under Test}
\acrodefplural{SUT}{Systems Under Test}
\acroindefinite{SUT}{an}{a}
\acro{LDCOF}{Local Density Cluster-Based Outlier Factor}
\acroindefinite{LDCOF}{an}{a}
\acro{DTW}{Dynamic Time Warping}
\acro{BMB}{BenchMark Bot}
\acro{ROS}{Robot Operating System}
\acro{NREC}{National Robotics Engineering Center}
\acroindefinite{NREC}{an}{a}
\acro{FDR}{False Detection Rate}
\acroindefinite{FDR}{an}{a}
\acro{UAS}{Unmanned Autonomous System}
\acro{ARS}{Autonomous and Robotics System}
\acro{ML}{Machine Learning}
\acro{IDS}{Intrusion Detection System}
\acro{CPS}{Cyber-Physical System}
\acro{DBI}{Dynamic Binary Instrumentation}
\acro{API}{Application Programming Interface}
\end{acronym}
\title{Using Dynamic Binary Instrumentation to \\
Detect Failures in Robotics Software}

\author{Deborah~S.~Katz,
        Christopher~S.~Timperley,
        and~Claire~Le~Goues% <-this % stops a space
\thanks{D.S. Katz performed this work when with the Computer Science Department, Carnegie Mellon University, Pittsburgh,
PA, 15213 USA e-mail: dskatz@gmail.com . She is currently affiliated with Seegrid.}% <-this % stops a space
\thanks{C. Le~Goues and C.S. Timperley are with the Institute for Software Research, Carnegie Mellon University.}% <-this % stops a space
%\thanks{Manuscript received \todo{date}; revised \todo{date}.}
}

\maketitle

\begin{abstract}
\acp{ARS}
are widespread, complex, and increasingly
coming into contact with the public.
Many of these systems are safety-critical, and it is vital to detect
software errors to protect against harm.

We propose a family of novel techniques to detect unusual program executions
and incorrect program behavior.
We model execution behavior by collecting low-level signals at 
run time and using those signals to build machine learning models.
These models can identify previously-unseen executions that are more likely
to exhibit errors.

We describe a tractable approach for collecting dynamic binary runtime signals
on \acp{ARS}, allowing the systems to absorb most of the overhead
from dynamic instrumentation.
The architecture of \acp{ARS} is particularly well-adapted to hiding the overhead
from instrumentation.

We demonstrate the efficacy of these approaches on \tool{ArduPilot} --- 
a popular open-source autopilot software system --- and \tool{Husky} 
--- an unmanned ground vehicle ---
in simulation.
We instrument executions to gather data from which
we build supervised machine learning models of executions
and evaluate the accuracy of these models.
We also analyze the amount of training data needed to develop models with
various degrees of accuracy, measure the overhead added to
executions that use the analysis tool, and analyze which runtime signals 
are most useful for detecting unusual behavior on the program under test.
In addition, we analyze the effects of timing delays on the functional
behavior of \acp{ARS}.
\end{abstract}


% \fi
\IEEEpeerreviewmaketitle

\section{Introduction}
\IEEEPARstart{A}{utonomous} and Robotics Systems (ARSs) are widespread, complex, and increasingly
coming into contact with the public.
Many of these systems are safety-critical, and it is vital to detect
software errors to protect against harm~\cite{FraadeBlanarMeasuring2018, HutchisonRobustness2018}.
Significant challenges hamper efforts to ensure end-to-end safety in autonomous
systems~\cite{KoopmanAutonomous2017}.  Such systems are often
inaccessible and resource constrained (in terms of both power and computing
resources), and typically make use of a mix of custom- and off-the-shelf
components from a diversity of suppliers~\cite{ForrestChallenges2014}.  Use of
proprietary components often 
means that source code is unavailable~\cite{SchulteEmbedded2013}. 
Similarly, it is typically difficult to
adequately model certain system elements, especially continuous
elements~\cite{SanwalVerification2013}, for the purpose of formal verification.
System correctness and safety thus
requires a diverse array of assurance and analysis techniques, and those
techniques typically cannot assume  the availability of either source code or
formal specifications for analysis.
 
However, robotics software also presents unique opportunities to leverage
techniques that are not applicable or practical in standard software
applications~\cite{SchulteEmbedded2013, ChengTaint2006}. In
particular, we observe that, as parts of
a real-time, distributed system, robotics system
components often experience significant idle time waiting for events from other
parts of the system or from the environment. 
This presents valuable unused capacity that can ``hide'' the overhead of
tools that are
impractical to use with CPU-, I/O-, or memory-bound software.
We demonstrate some of this capacity experimentally.

With this opportunity in mind, 
our key insight is that \emph{Dynamic runtime characteristics can
tell us about program behavior,
from which we can build a machine-learning model of expected behavior.}
That is, by observing and measuring low-level execution signals such as
number of
machine instructions executed, we can effectively characterize program behavior
over many executions.  By combining these signals using machine learning, we can
produce models that determine whether the current execution appears normal or
anomalous, in terms of deviation from  established
patterns,~\cite{EnglerDeviant2001}. These models can be used to predict 
whether new executions can
be categorized as behaving as intended, or exhibiting errors.
Our further insight is that \emph{the architecture of \acp{ARS} makes this approach tractable despite timing-sensitivity}.

Marrying these two observations, in this paper, we present an approach we call
\thename for detecting errors in robotics systems. 
\thename uses \ac{DBI} to collect rich,
low-level information about program behavior.
We use machine learning to
analyze and combine the signals into predictive models that identify whether an
execution is nominal or anomalous. 

At a high-level, \ac{DBI} inserts code
that analyzes a subject program while it executes. 
Although it is typically too
heavy-weight an approach to be practical in many
circumstances~\cite{SchulteEmbedded2013, ChengTaint2006}, we find that robotics systems are
well-suited to an approach that uses \ac{DBI} appropriately, because they can absorb 
overhead in component idle time.
We find this to be the case, even in cases when the timing of control loops has been tuned to avoid waste.
Despite this opportunity, however, significant effort is required to still keep
overhead tractable. 
Because they are real-time systems, robotics systems are sensitive to
timing, sequencing, and deadline issues that can be affected by excessive instrumentation
overhead. 

We therefore develop a custom \ac{DBI} tool we call \dbitool using the \tool{Valgrind}
platform~\cite{NethercoteValgrind2007, NethercoteValgrind2004} to efficiently collect selected runtime information that summarizes key
characteristics of an execution's behavior, such as the number of instructions
executed or the maximum address of a memory load~\cite{KatzDetecting2020}.

% CLG I kind of like the below sentence but can't really figure out how to
% incorporate it.
%One challenge in implementing a tool based on dynamic binary instrumentation
%is determining which pieces of information to collect, with an eye to
%keeping overhead low.

\thename is applicable either at development- or run-time. 
At development time, in simulation or field testing, our techniques can detect
unusual executions, providing the opportunity to repair any underlying faults
in the software before deployment.
Many errors in robotics software can be
replicated in software simulation without full environmental
replication~\cite{SotiropoulosNavigationBugs2017, TimperleyCrashing2018}.
Discovering robotics faults early in simulation can reduce \ac{QA} costs as well as
the cost of expensive field-test failures~\cite{WilliamsonCost2008}. 
At run time, our techniques can detect unusual elements in executions before
an error results in an outwardly-observable failure mode, providing the
opportunity to stop the system or put it into a fail-safe mode.
There are many use cases for alerts about unusual software behavior, intended or
otherwise, including 
putting a critical system into a failsafe mode while a potential
error is investigated~\cite{ForrestChallenges2014},
robustness testing~\cite{HutchisonRobustness2018, KroppBallista1998},
and intrusion detection~\cite{DenningIntrusion1987, WagnerMimicry2002, CylanceopticsBrief2018}.
Additionally, \thename is language-independent and operates at the
binary execution level, without a need for source code, fitting the needs
for analysis of many systems in the \ac{ARS} domain.

Our work is complementary to previous work that has attempted to detect 
faulty program executions,
some of which does so by establishing patterns of program behavior.
%Statistical debugging~\cite{ZhengStatistical2006, ZhengSampled2004, LiblitScalable2005, LiblitIsolation2003}
Automated invariant detection techniques~\cite{EnglerDeviant2001, ErnstDaikon2001, ErnstDaikon2007, HangalTracking2002, HangalIodine2005, PerkinsClearview2009}
and statistical debugging techniques~\cite{ZhengStatistical2006, ZhengSampled2004, LiblitScalable2005, LiblitIsolation2003} automatically identify properties 
that hold true over all correct executions of a program and identify bugs 
via violation of those invariants.
However, such techniques typically require
source code, can struggle to scale, or have other limitations which reduce
their usefulness in complex autonomous systems.
Formal verification of cyber-physical systems is powerful,
but its practical application to entire real systems is limited.
Models can include assumptions that do not necessarily hold true,
have inadequate modeling of the physical world, and can require an 
extraordinary investment of time and human
effort~\cite{SanwalVerification2013, ZhengVerification2015, ZhengState2017,
KleinVerification2010, KleinVerification2014}.
Related techniques are given a more complete treatment in Section~\ref{sec:related-work}. 

We present \thename as well-suited to robotics applications 
and test it on the \tool{ArduPilot}
software, which provides auto-pilot systems for various autonomous vehicles, as
described in Section~\ref{sec:background-ardupilot}. 
We evaluate it on indicative executions of \tool{ArduPilot},
assessing the accuracy of the models' output and evaluating
training time and overhead.
We show that \thename is highly accurate at detecting executions that exhibit
errors and detects those errors with minimal overhead.
We further evaluate the extent to which overhead impacts observable robot
execution.

The main contributions of this work are as follows:

\begin{itemize}
\item A tractable approach for using dynamic binary instrumentation
	as an analysis tool for robotics systems.
\item  An approach, \thename, for using dynamic binary runtime signals as input to
  supervised machine learning techniques to detect unusual execution
  behavior.
\item An evaluation of \thename on simulations of the \tool{ArduPilot}
 robotics software, to detect executions that exhibit unusual behavior.
\item An evaluation of the amount of training data and overhead needed to use \thename.
\item An analysis of the dynamic binary runtime signals most useful
 for detecting unusual behavior in \tool{ArduPilot}.
%\item An analysis of changes to the overhead of the dynamic binary analysis
% tools on \tool{ArduPilot} when data collection is limited to a reduced set
% of useful signals.
\item An analysis of the effects of delays on the execution behavior of %\tool{ArduPilot} and 
\tool{Husky}, in support of the use of dynamic instrumentation as a tool for analyzing timing-critical systems.
\end{itemize}

\section{Background and Motivation}

This section provides background for our work, including laying out
details of the programs
on which we evaluate our techniques: \tool{ArduPilot} and \tool{Husky}.
It also walks through a motivating example, based on one of the experimental
scenarios we evaluate for \tool{ArduPilot},
and provides an introduction to \acf{DBI}.

\subsection{The Subject Systems}
\label{sec:background-ardupilot}

\paragraph{\tool{ArduPilot}}
We run the majority of our experiments on the \tool{ArduPilot} system.\footnote{\url{http://ardupilot.org}}
This is an open-source project, written in C++, with autopilot systems
that can be used with various types of autonomous vehicles.
It runs a control loop architecture.
\tool{ArduPilot} is very popular with hobbyists, professionals, educators, and researchers
and has approximately 678 thousand lines of code and over 50 thousand
commits in its GitHub repository.\footnote{https://github.com/ArduPilot/ardupilot}

\tool{ArduPilot} provides a rich ground on which to test robotics systems.
It is sufficiently complex to be useful in the real-world.
There is a wealth of information about bugs encountered in real world usage,
both in the version-control history and in the academic
literature~\cite{TimperleyCrashing2018}.
We evaluate \thename on \tool{ArduPilot} in simulation, using the included
software-in-the-loop simulator.
We use a customized test harness that enables coordinated control over
simulations.

For explanation and clarity, we present a simplified overview of relevant features of \tool{ArduPilot}'s operation:
\tool{ArduPilot} executes startup commands to prepare the controller
and start the autonomous vehicle.
Then, if the system is in autopilot mode, 
the controller operates on a control loop.
In practice, this means that the controller becomes ready to receive
commands, one at a time, and executes them autonomously.

\paragraph{\tool{Husky}}
We evaluate the timing experiments on \tool{Husky}.

The
\tool{Husky} unmanned ground vehicle by Clearpath
Robotics\footnote{\url{https://clearpathrobotics.com/husky-unmanned-ground-vehicle-robot/}}
is a real world robot with an extensive simulation infrastructure.
It is rugged, designed to be deployed in uneven terrain, and it is capable of 
carrying and integrating with a variety of input sources (sensors) and actuators. 
Husky is popular among researchers for its straightforward design and real world 
usage history.

\subsection{Motivating Example}
\label{sec:background-motivating}
As a motivating example, we present a memory corruption bug in
\tool{ArduPilot}, specifically the \tool{ArduCopter} software for autonomous
control of airborne vehicles.
% \todo{add some source code here? based on CT's comment}
The example we present here is based on a fault we seeded in early
experiments with \tool{ArduCopter}.

For each valid command received, the controller executes corresponding
code, as defined in \tool{commands\_logic.cpp}.
In this example, the code corresponding to a particular command ---
\texttt{MAV\_CMD\_NAV\_CONTINUE\_AND\_CHANGE\_ALT} ---
in
\tool{commands\_logic.cpp} is seeded with a buffer
overflow.
This does not cause an externally-noticeable issue until the code corresponding
to a different command --- \texttt{MAV\_CMD\_NAV\_PAYLOAD\_PLACE} --- attempts to read
the data in the location that the overflow had written to.

At a low-level, looking at the pattern of memory writes can reveal
the difference between the executions in which memory is corrupted and
the executions in which it is not.
For example, in the execution of the code corresponding to the
command in which the memory is corrupted, there will be writes to memory
addresses that are not usually written to in that part of the program.
There may also be an unusually high number of memory writes.
In addition, the difference may show up in other low-level indicators
that we do not intuitively associate with buffer overflows.

\thename uses machine learning techniques to look at all of the low level data collected by \dbitool together.
This gives us the benefit of not needing to know in advance which of
the indicators may be relevant to any given bug.

In addition, monitoring low-level signals has the potential to
detect this issue before
there is an externally-visible problem by analyzing
data collected after the buffer overflow is written but
before the program gets to the code in which the corrupted memory is accessed.
If we collect signals through the execution of the command
\texttt{MAV\_CMD\_NAV\_CONTINUE\_AND\_CHANGE\_ALT} but analyze them before
the execution of \texttt{MAV\_CMD\_NAV\_PAYLOAD\_PLACE},
the approach has the potential to detect the problem that has already been set up
but has not yet caused any externally-observable effects.
This flexibility can allow \thename to detect errors before they manifest
as user-observable crashes.

\subsection{Dynamic Binary Instrumentation}
\label{sec:background-dbi}
\acf{DBI} works by analyzing a \ac{SUT} at execution time.
The \ac{DBI} analysis tool inserts code --- \emph{instrumentation} ---
that analyzes the \ac{SUT}, to be run while the \ac{SUT} runs.
Because \ac{DBI} works at runtime, it can encompass
any code called by the subject program, whether it be within the original
program, in a library, or elsewhere~\cite{NethercoteValgrind2004,NethercoteValgrind2007,LukPin2005}.

As a dynamic binary instrumentation framework,
\tool{Valgrind}\footnote{\url{http://valgrind.org/}} allows
tools based on the platform to record data about low-level
events that take place during a program's execution.
Tools based on \tool{Valgrind} can have significant overhead,
although optimizations can reduce that overhead.
For example, running the example tool, \tool{Lackey}, and memory
checking tool, \tool{Memcheck}, that are
distributed with \tool{Valgrind} on a simple command in a basic \tool{Linux} utility 
results in a 404x and 383x overhead, respectively, as measured by execution time.
Our custom tool based on \tool{Valgrind} incurs 186x overhead on this same
command.
The overhead for these tools on \tool{ArduPilot} is significantly
smaller, as discussed in Sections~\ref{sec:evaluation-overhead}
and \ref{sec:evaluation-delays},
inspiring the use of \ac{DBI} for robotics systems.
We describe engineering considerations that make it possible to use
a \tool{Valgrind} tool on a robotics system in Section~\ref{sec:discussion-dbi-overhead}.

\section{Approach}
\label{sec:approach}

\begin{figure}
%\centering
  \adjustbox{max size={0.95\linewidth}{0.95\textheight}}%
            {\includegraphics[scale=0.95]{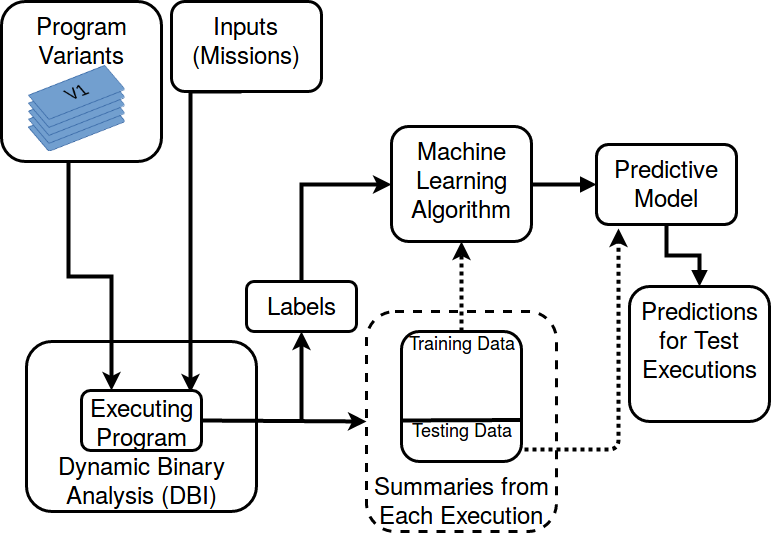}}
  \caption{\label{fig:supervised-arch}General Overview of the Supervised Learning Approach}
\end{figure}

In this section, we describe our approach for identifying software errors
using supervised learning models built over low-level dynamic signals.
We envision this technique --- \thename --- to be used either at development time
or during field testing.
At development time, \thename can be used on robot executions in simulation,
to detect potential errors before deployment on real hardware.
In deployed hardware during field testing, \thename can be used to issue alerts 
regarding probable software errors,
which can be addressed by putting the software or system into a fail safe
mode while the problem is addressed.

We explain \thename with reference to Figure~\ref{fig:supervised-arch},
which lays out the architecture of the approach.
We begin by generating a corpus of data that includes software
executions that exhibit an error and executions that do not.
Details about options for generating this corpus are in 
Section~\ref{sec:approach-corpus}.
For the purpose of the experiments in this paper, we begin generating the
corpus with with a base version of the \ac{SUT}.
From the base version, we generate mutant program variants using a
systematic mutation approach.
We also have a set of program inputs, which we can think of as
defect-revealing inputs and non-defect-revealing inputs.
As illustrated in Figure~\ref{fig:supervised-arch}, each program variant
is executed with each input.

We monitor these executions with a custom dynamic binary analysis (DBI)
tool --- \dbitool --- as described in Section~\ref{sec:approach-dbi}.
The tool produces a \emph{summary} of each execution, which consists
of a set of counts and extremes of various events that occur at runtime.
We call these pieces of data \emph{signals}.

When each execution runs with a given input, we assign it a label of
\emph{pass} or \emph{fail}, corresponding to whether the execution
exhibited expected behavior.
These labels are necessary as input to the supervised learning algorithms
we use.
We also use the labels to evaluate the success of our techniques.
Using labeled data as input is a common technique, and labels for training
data can be obtained in many ways.

We use the sets of signals as input to a \emph{supervised machine learning} algorithm.
This algorithm generates a \emph{predictive model}.
The predictive model takes, as input, \emph{test data} ---
signals from executions that were
not included in the data used to build the model.
The model outputs a boolean prediction in the form of a 1 or 0
 as to whether each of those sets of signals
corresponds to an execution that exhibits an error.\footnote{The word,
``prediction,'' and related terms are terms of art in machine learning.
Here, they refer to any time a machine learning algorithm makes a determination
on previously-unseen data.}

\subsection{Collecting Signals with \acl{DBI}}
\label{sec:approach-dbi}
To record low-level information about program executions,
we developed a custom tool --- \dbitool --- based on the \tool{Valgrind}
framework,\footnote{\url{http://valgrind.org/}} 
version~3.14.
As a \ac{DBI} framework, \tool{Valgrind} allows
tools based on the platform to record data about low-level
events that take place during a program's execution.
Examples of such low-level events are the execution of a single instruction
or a single load from memory into a register.

We use \ac{DBI} to collect \emph{signals} that summarize the behavior of the
execution monitored.
The signals are chosen to be potentially informative
while being easy to collect within the \tool{Valgrind} framework and
incurring very little
calculation or data storage overhead at runtime.
We discuss these choices in Section~\ref{sec:discussion-dbi-overhead}.
These counts result in a \emph{summary} of each execution, consisting of a list
of numbers, corresponding in order to the set of properties measured.

Our \ac{DBI} tool outputs these summaries at intervals throughout the execution.
Each signal is measured cumulatively from the beginning of the execution.
By analyzing the summaries of signals at each interval, our technique 
can detect executions that exhibit defects that occur in the execution time
covered by each interval.
As discussed in Section~\ref{sec:background-motivating}, unusual low-level
behavior can occur well before an error is outwardly-visible.

We collect the same measurements for each summary, so the summary
can be treated as a feature vector, suitable for input
into machine learning algorithms.
We aggregate the feature vectors over many executions into
two-dimensional matrices,
representing the overall data set.

In all, this customized tool outputs 26 signals.
These signals include: counts of machine instructions executed,
basic blocks entered and exited,
load and store instructions executed, and
events tracked internally by \tool{Valgrind}; and
minima, maxima, and ranges of addresses of
machine instructions executed and
data loaded and stored. 
It is possible to obtain accurate results with a subset of these signals,
as discussed in Section~\ref{sec:evaluation-features}.
Our experiments use all 26 of these signals unless otherwise specifically
mentioned. %\footnote{Link to code and details of signals to
%be provided for the camera-ready. \dsk{probably need to get this together}}

\subsection{Minimizing Overhead in Dynamic Binary Instrumentation}
\label{sec:discussion-dbi-overhead}

%\todo{CLG says: this subsection would benefit from significantly more detail.
%  Fewer ``for examples'' and more concrete lists of the signals you're
%  collecting and how.  Imagine a reader wants to do DBI on a robot based on this
%  paper.  I'm always yelling about there being too much implementation detail;
%  this is the one case where I think the paper would benefit from significantly
%  more of it!}

One key difficulty in using a dynamic binary instrumentation approach
with a timing-sensitive system is that the overhead of collecting the information
changes the timing in the program execution.
%These timing changes can change the program's control flow by, for example,
%causing various timeouts to trigger or causing events to happen in an order
%that the program does not expect.
For example, \tool{ArduPilot} has several timeout windows during which
it expects certain events to happen; if they do not, the system aborts.
We used several approaches in tool design to tractably reduce overhead:

\vspace{1ex}
\noindent\textbf{Optimize to the basic block level whenever possible.}
Minimizing runtime interruptions is key to reducing instrumentation overhead.
One established approach is to restrict interruptions to once per basic block,
rather than every instruction.
At the basic block level, it is possible to collect a summary of the events that
occur within that block.
However, not all information we wish to collect is available at the basic block
level.
For example, addresses that are computed at runtime may not be available.
We, therefore, optimize instrumentation to the basic block level when
all of the information we want is available at that level,
such as when the block does not rely on any addresses computed at runtime.
Otherwise, we instrument instructions within the block individually.

\vspace{1ex}
\noindent\textbf{Eliminate excess data storage at instrumentation site.}
The amount of data associated with instrumentation (stored and retrieved at
runtime) can drastically influence the overhead.
We experimented with an approach that collected richer data by storing histograms
of many data points.
These included instruction addresses, memory addresses,
and data values, along with more esoteric data; 
we used the histograms to calculate summary signals, such as the most frequent value.
However, the memory usage and library calls involved in storage and updating the
histograms slowed the instrumentation dramatically.
Ultimately, simple tallies that involved a minimum of data storage and operations
at runtime proved more tractable.

\vspace{1ex}
\noindent\textbf{Avoid high-overhead \ac{API} calls and computation at
  instrumentation sites.}
Because we must add many instrumentation sites, the overhead at each site must
be minimal. 
Fortunately, not all low-level events are equally expensive (in time and data) for \tool{Valgrind} to measure. 
For example, counts of information that require additional calls into the
\tool{Valgrind} \ac{API}, such as some branch prediction information, are more expensive
to collect than counts of information already available
to the tool at an instrumentation site, such as the address of the current
instruction.
We restrict the events we measure to those that can be measured
with a minimum of added overhead.
For example, we do collect the address and type of each instruction.
We do not collect data related to the order of memory events that
would be useful if we were explicitly tracking memory exceptions.

\vspace{1ex}
Overall, instead of tracking high-overhead information,
we take best advantage of each instrumentation
site by collecting as much information as possible that is available 
with minimal calculation.
To do this, we build \dbitool on top of \tool{Valgrind}'s
\tool{Nullgrind} tool, a tool designed to do nothing.
We collect data on events tracked by \tool{Valgrind} internally because
those data are readily available.
We keep a count of various types of instructions instrumented, even when
those types of instructions are internal \tool{Valgrind} bookkeeping
instructions that do not correspond to machine instructions.
We track these because it does not cost any additional overhead,
and they may correspond to useful information.

\subsection{Corpus Generation}
\label{sec:approach-corpus}

Using \thename requires a corpus of labeled data for supervised learning,
which we discuss in Section~\ref{sec:approach-supervised}.
Generating this corpus of data requires a \ac{SUT} that can be executed
under \ac{DBI} and a way of labeling each execution
as exhibiting intended or unintended behavior.
The corpus must contain at least some data corresponding to executions
with intended behavior and some data corresponding to executions
with unintended behavior.
Examples from both classes are needed for the two-class supervised
learning algorithms used as a part of \thename.

There are many possible ways to generate a set of executions of the \ac{SUT}
that includes both well-behaved and misbehaving executions.
Some \acp{SUT} contain faults in their code as written
and, when executed with an appropriate input, will exhibit errors.
Another way to obtain executions that exhibit errors is to seed faults into
the underlying program code.
Again, with an appropriate input, the execution will exhibit an error.
Ways to seed faults include mutating
the programs using techniques common in mutation testing or
seeding a known error into the code.
One way to seed a realistic error is to use the edit history of a program's
source code to find a change that repaired a bug, then re-introduce the
corresponding `buggy' code into the \ac{SUT}~\cite{TimperleyMutation2017}.
Note that obtaining faulty executions by seeding faults does not limit
\thename to only programs for which source code exists because
faults can be seeded at any level, including in machine code.
An additional way to obtain executions that exhibit errors is to
include several versions from the version history of the same program,
as long as those versions are sufficiently similar to one another to compare executions.

For our experiments, we generate the corpus by taking the base program
as ArduPilot 3.6.7.
We generate mutants using established techniques in mutation testing,
with a primitive set of source code mutation operators used in prior
studies~\cite{ChenLearning2018}.
We construct each mutant by applying a single mutation operator to a location
in a core source file.
We construct one such mutant for every combination of applicable location and 
mutation operator in several core source files.
We use BugZoo~\cite{TimperleyBugZooPoster2018} to create an ephemeral Docker container 
for each mutant under test.
We execute each mutant with a suite of three inputs,
which in the case of \tool{ArduPilot} are known as \emph{missions},
and discard data for any mutants that fail to compile and for any executions that
crash immediately when executed.
This results in \numexecutions execution traces.
We then use a simple oracle to classify the traces as
\emph{correct (pass)} or
\emph{erroneous (fail)} based on the simulated physical location of the robot during
the mission.
Following this classification, we are left with \numunbalancedcorrect correct
and \numunbalancederrors erroneous traces.
To avoid issues of class balance which can bias machine learning training,
we balance all data by duplication of
pseudo-randomly chosen data points in the minority class.
We choose to balance by duplication because it does not reduce the size
of the data set; our preliminary experiments show comparable results
with balancing by deletion, when the data set is large enough.
We balance the data separately for each model we build,
and balance each fold separately for K-fold cross validation,
to avoid the possibility of duplication resulting in the same summary
appearing in the training corpus and test inputs.
%After balancing the data, we are left with \numbalancedcorrect correct
%and \numbalancederrors erroneous traces.

\subsection{Supervised Learning}
\label{sec:approach-supervised}

Supervised learning is a technique for developing machine learning
models to classify data.
We use two-class supervised learning, which means that the data falls
into two categories.
In this case, the categories are \emph{pass} (0), which corresponds
to executions that exhibit intended behavior, and \emph{fail} (1), which
corresponds to executions that exhibit unintended behavior.
A supervised learning algorithm takes, as input, \emph{labeled} data
in the form of \emph{feature vectors} ---
which in this case are the \emph{summaries} output by our custom
DBI tool \dbitool --- each of which has a corresponding
0 or 1, indicating the category to which the corresponding data belongs.
That is, data that came from executions that were known to exhibit an error
have the label 1, while the rest of the data came from executions that
were not known to exhibit an error.
We used a simple oracle of our own design to generate the labels,
otherwise known as \emph{ground truth}.
This input data is known as \emph{training data}.
The labels for training data could come from the use of test cases, human observation
of execution behavior, or other sources of knowledge of whether
an execution is correct.

The supervised learning algorithm outputs a \emph{classifier}.
The classifier takes, as input, \emph{test data} which consists 
of feature vectors -- summaries -- 
in the same form as those used to train the classifier but that were not used to train the classifier.
The test data does not include labels.
The classifier outputs \emph{predictions} -- determinations as to which
class each data point belongs to.
The term \emph{predictions} is a term of art in machine learning
that refers to the output of a classifier;
it does not necessarily refer to future events.
We assess the accuracy of the classifier by comparing the predictions
to the known ground truth, which we obtain from our simple oracle.
We discuss accuracy metrics in Section~\ref{sec:experimental-metrics}.

For supervised learning, we use use out-of-the box algorithms from
Scikit Learn, version 0.15.2.\footnote{\url{http://scikit-learn.org/}}
Specifically, we use the Decision Tree classifier
available from Scikit Learn.
Based on an informal survey of the options available from Scikit Learn,
we found that the Decision Tree classifier performs similarly to
or better than other algorithms when used as a part of this technique,
across a wide range of programs and data.
We use the default parameters of the algorithm.

\subsection{Approach and Methodology for Timing Experiments}
\label{sec:approach-timing}

To evaluate the extent to which timing delays deform the observable execution of an
\ac{ARS}, we use a different set of experiments, in which we insert artificial timing delays in a controlled manner.

For a given robot, we establish a set of commands, called a \emph{mission}.
For the purposes of these experiments, each mission is represented as a series 
of destinations in three dimensional space (two dimensional space for robots that 
move in only two dimensions), 
with the final destination being a return to the first destination.
We create a series of missions within a simulation environment for each robot.

The experiments consist of running two types of executions: \emph{nominal baseline}
executions in which the system is run without modifications and \emph{experimental}
executions in which the system is run with artificially-inserted timing delays.

\paragraph{Nominal Baseline Executions}
To establish a \emph{nominal baseline} --- a baseline for how a robot behaves under 
normal conditions, without 
any artificially-inserted delays --- 
we run each unmodified \ac{ARS} repeatedly on each of its missions.
%For the \tool{ArduPilot} experiments, I run the unmodified system on each mission 20 
%times and for the \tool{ROS} experiments,
%I run the unmodified system on each mission 200 times.
%\todo{CHECK THESE NUMBERS}
%\todo{these numbers don't belong here -- use before def. where do they belong?}

%From the nominal executions, I use statistical methods, described below in
%Section~\todo{cross-ref}, to choose a representative normal execution against
%which to compare the artificially-delayed executions.
%Actually, you didn't decide to do it this way -- don't say so.

The executions to establish a nominal baseline serve several purposes in these experiments.
First, the nominal executions establish a baseline for how often the unmodified
\ac{ARS} fails.
\acp{ARS} often behave in a nondeterministic manner, even in simulation.
%Factors that can affect the nondeterminism include variations in the order in which
%messages are received, sensor noise, perception systems' interpretation of the
%sensor data, autopilot systems, and obstacles in the environment.
There can be nominal (unmodified) executions that fail in significant ways, such
as failing to reach one or more waypoints; getting ``stuck'' and discontinuing
attempts to follow the mission (e.g., when the perception system cannot determine
the robot's location); software crashes; or liveness failures (e.g., hitting a
timeout).
The number of failures in the nominal, unmodified system establishes a point
of comparison by which to measure the
failure rate of the artificially-modified system.
As a generalized approximate metric of these failures,
we use the percent of executions
in which the robot never reaches the final waypoint.
This metric allows comparison between the failure rate in nominal executions
and the failure rate in the modified system.

Second, the nominal executions establish a representative trajectory and other
execution characteristics against which the characteristics of modified executions
can be compared.
Other potential characteristics of interest include the time taken for completion of 
the mission and the rate at which messages are sent on various topics.
% \todo{finish this paragraph}

Third, the nominal executions establish the range of variation in nominal 
trajectories and other execution characteristics. As mentioned above, 
there is significant nondeterminism in the observed behavior of simulated robots,
even when unmodified.
Establishing the range in the nominal executions provides a basis to tell when the
modified, experimental executions are within the range of nominal behavior or
outside of it.

\paragraph{Experimental Executions}
For the \emph{experimental} executions, we add controlled artificial delays to the 
execution of the \ac{ARS} code.
The experimental parameters include: the points within the program at which these delays are inserted, the number of insertion points, and the length of delays.
The method of inserting delays is set out below.

Experimental executions are evaluated against 
the nominal baseline and against
the set of waypoints that the simulated robot is directed to reach. 
%\todo{add more about the experimental executions and how they're compared to the nominal baseline and why}

\subsubsection{Subject Systems}

In addition to the \tool{ArduCopter} \emph{\tool{ArduPilot}} system, which features heavily in the other experiments
in this paper, we also evaluate on a \ac{ROS}-based system: \emph{\tool{Husky}},
to demonstrate the generalization of the timing absorption to other \ac{ARS}.

\subsubsection{Method of Inserting Delays}
\label{sec:timing-methodology-delays}
This subsection explains how artificial delays are inserted for the experimental 
executions.

\paragraph{\tool{ArduCopter}}
For the \tool{ArduCopter} experiments, the artificial delays are introduced
by modifying source code in \tool{C++} and inserting \tool{sleep} statements.
We identify the program point before each return statement
in all \tool{.cpp} files in the
\tool{ArduPilot/ArduCopter} source code directory.
Each of these program points was a possible location to insert a delay.
The choice of whether to insert a delay at each point was determined probabilistically, with a weighted coin flip.
Different modified versions of the code were created,~\footnote{We used the tool \tool{Comby} for these program transformations. \url{https://comby.dev/}} each of which had
(a) a fixed coin flip weight and
(b) fixed delay amount added at each delay location.
The weights for the weighted coin flip ranged from 0.1 to 1.0, with 1.0
meaning a delay was inserted before every return statement,
and the length of each delay ranged from 0.001953125 seconds to 8 seconds,
with delay lengths chosen as powers of 2.

\paragraph{\tool{Husky}}
We conducted similar experiments on \tool{Husky}, which is a system based in
\tool{\ac{ROS}}.
For the \tool{\ac{ROS}} experiments, the artificial delays are introduced at 
communications
barriers on \tool{\ac{ROS}} topics, taking advantage of the architecture of 
\tool{ROS}-based systems.

\begin{figure*}
  \centering
  \adjustbox{max size={0.65\linewidth}{0.65\textheight}}%
            {\includegraphics{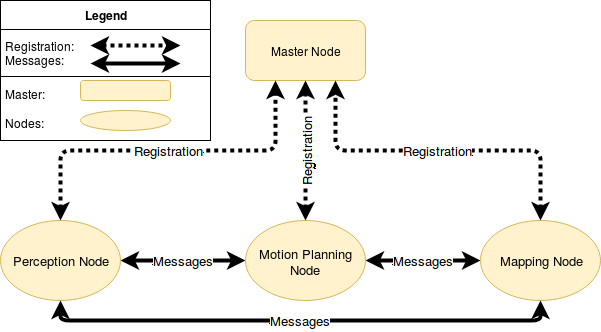}}
  \caption{\label{fig:ros-arch-2}Simplified \tool{ROS} Architecture}
\end{figure*}

To give a simplified overview of the architecture of \tool{\ac{ROS}}-based systems as
they relate to these inserted delays,
these systems consist of various nodes that communicate with each other by
sending messages over a bus, as shown in Figure~\ref{fig:ros-arch-2}.

%\todo{Edit the ROS architecture diagram to show topics and message types traveling
%along those topics.}
A publish-subscribe system determines which nodes receive which messages.
A node can publish messages to a \emph{topic}.
To receive those messages, another node subscribes to the same \emph{topic}.
Generally, each topic only accepts messages of one type.
\tool{\ac{ROS}} makes it easy to query a running system to find out configuration
information
such as (a) the topics in that system; (b) the type of messages published to each
topic; (c) the node or nodes that publish to a given topic; and (d) the node
or nodes that subscribe to the given topic.
This information makes it easy to infer certain properties about the relationships
among nodes and the purposes of particular messages.
We use this information to choose the topics to which we add artificial delays.
For example, in \tool{Husky}, we run a set of experiments that delay each topic published
by or subscribed to by the \tool{/move\_base/} navigation node.
We make this choice because navigation is a vital function, and we expect disruptions
in navigation to have an effect on robot behavior.
By contrast, we do not conduct experiments in which we delay the topics related to
displaying logging messages, as we do not expect these delays to affect the robot's functional behavior.

Having chosen a \tool{topic} or \tool{topics} to delay on a particular \tool{\ac{ROS}}
system for a particular set of experiments, we insert delays on these topics by
intercepting messages using topic renaming.
\tool{\ac{ROS}} allows configuration of nodes such that topics can be renamed.
For example, if Node A is originally designed to publish a topic 
named \tool{/a\_very\_good\_topic}, we can change the system's configuration so that 
when published,
the topic is known in the namespace as something else, such 
as \tool{/a\_very\_good\_topic\_intercepted}.
Because the topic now has a different name than other components in the system
expect, the node or
nodes that would have originally subscribed to the topic will not receive
the messages published on the new topic.
However, the delay infrastructure includes an additional node
to be included with the \tool{\ac{ROS}} system.
This node reads each message on a given topic, in this
case \tool{/a\_very\_good\_topic\_intercepted}. 
It then waits for the designated amount of time and then republishes the same
message on the topic that was originally expected: \tool{/a\_very\_good\_topic}.
The nodes that originally expected the messages on this topic from Node A now receive
the same messages from the delay node.
%\todo{include diagram if you have time}

Delays range in length from 0.00390625 seconds to 1 second, chosen as powers
of two, and 
were inserted for every message in a topic. 
This range was chosen after a parameter sweep revealed that they
result in a representative range of behaviors.

\section{Evaluation}
\label{sec:experimental}
Using \tool{ArduPilot} 3.6.7 as a case study system, 
we evaluate claims in two major areas: (A) Performance and accuracy, and
(A) Effects of overhead.
we ask the following
research questions.

\paragraph*{Performance and Accuracy}
\begin{itemize}

\item RQ1: To what extent can \thename accurately detect executions that
 exhibit errors?

\item RQ2: How does the volume of training data used to build the model affect
 the accuracy of the model in detecting executions that exhibit errors?

\item RQ3: To what extent can \thename accurately detect executions that
 exhibit errors before reaching the end of the execution?
\end{itemize}

\paragraph*{Effects of Overhead}
\begin{itemize}

\item RQ4: How much overhead is introduced by \ac{DBI}? 

\item RQ5: What effects do delays have on observable robot performance? 
%\dsk{Merge this with the overhead timing chapter.}

\item RQ6: To what extent does feature reduction affect overhead and
 accuracy?

%\item RQ6: Can a model trained on executions of a prior release of the
% software under test identify errors in a subsequent release of the
% software?
\end{itemize}

We run all experiments on an Ubuntu 16.04 virtual machine
with 4 virtual cores and and 11GB RAM, 
running on a physical machine with an Intel(R) Xeon(R) E5620 CPU (2.40GHz).

\label{sec:experimental-metrics}
We assess the performance of the models produced by
\thename with respect
to their ability to correctly and generally label a set of traces.
Given a set of test data, known labels, and
the classifier's predictions on the test data, the predictions can be evaluated
into true and false positives and negatives (\textit{TP},
\textit{FP}, \textit{TN}, and \textit{FN}). Subsequent metrics include:

\begin{itemize}
\item \textbf{Accuracy} \textit{(Acc)} The portion of samples whose predicted labels match
the ground-truth labels: $(TP + TN) / (TP + FP + TN + FN)$.

\item \textbf{Precision}  \textit{(Prec)} The ratio of returned labels that are correct:
$TP / (TP + FP)$.

\item \textbf{Recall} \textit{(Rec)} The ratio of true labels that are returned:
$TP / (TP + FN)$.

\item \textbf{F-Score} \textit{(F)} The harmonic mean of precision and recall, 
which guards against trivially maximizing precision or recall by 
predicting the labels to be all negative or all positive. 
Calculated as: $2*((Prec * Rec)/(Prec + Rec)) $.
\end{itemize}

% Does this belong here? How is the seed used?
% We also use a fixed random seed for reproducibility.
Unless otherwise stated,
we use K-fold cross-validation,
with K=10 for all sample sizes greater than or equal to 100.
For smaller sample sizes, we use the largest K that ensures each fold has
at least 10 points.
We report the arithmetic mean across all folds.

\subsection{RQ1: Error Detection}
\label{sec:evaluation-rq1}
We answer RQ1:
To what extent can \thename accurately detect executions that
exhibit errors?

Recall that our overall goal is to determine whether \thename --- an
approach that combines dynamic-binary-instrumentation with
machine-learning analysis --- is useful in detecting behavior that
exhibits defects in software.
This question evaluates the overall suitability of this approach
to detect errors in completed program executions.
To answer this question, we take the approach described in
Section~\ref{sec:approach} and evaluate it on the data for all
executions in our corpus, as described in Section~\ref{sec:approach-corpus}.
For this question, we use the cumulative signals recorded at the end of
each execution.

We measure Accuracy, Precision, Recall, and F-Score, as described in
Section~\ref{sec:approach-supervised}.
We report the means across 10-fold cross-validation.
For each metric, higher is better and means that the technique's
determinations are more accurate.

Table~\ref{table:supervised-total} shows results. 
We find that we can build a strong classifier that detects errors in
complete \tool{ArduPilot} executions.
The precision is nearly perfect, which means that there are very few
false positives.
This classifier works across different defects and missions, supporting
its generality.

%\todo{make this explanation more enlightening}

\begin{table}[t]
\centering
\caption{Accuracy Metrics for Supervised Learning using Data at the
End of Execution
\label{table:supervised-total}}
%\begin{tabular}{r | r r r r | r | r}
\begin{tabular}{ r r r r | r}
\toprule
%Ins. & Mean & Mean & Mean & Mean & Exec. & Num.\\
%Exec. & Acc. & Prec. & Rec. & F-Score & Time (secs) & Samples \\

Mean & Mean & Mean & Mean & Num.\\
Acc. & Prec. & Rec. & F-Score & Samples \\

\midrule
0.95 & 0.99 & 0.90 & 0.94 & \numexecutions \\

\bottomrule
\end{tabular}

\end{table}

%Recall that our overall goal is <goal>.  This question
%evaluates <subgoal> of <goal>.
%To answer this question, we take <artifact/analysis> described in Section X
%and <run/execute> it on <corpus/data/subset thereof>
%as described in Section Y.
%We measure <metric>. <higher/lower> is better, and means <English prose>.
%<optional additional details specific to this experiment (or recall, for 
%these conditions, things we said before>
%Figure A shows results. <How to interpret data presented, e.g., the X axis is
%<X>, the Y axis is <Y>><special normalization or whatever>. 
%Note that: <interpretation>
%
%* We ask RQ1 because...
%* To answer RQ1 we...
%* We find that...
%* Summary of finding!

\subsubsection{Using a Prior Release to Detect Errors on a Subsequent Release}
\label{sec:evaluation-rq1-release}
We ask a related question:
 Can a model trained on executions of a prior release of the
 software under test identify errors in a subsequent release of the
 software?

We answer this question by training a model on a set of executions on 
\tool{ArduPilot} version 3.6.6
and testing it on executions from version 3.6.7.
The model trained on \tool{ArduPilot} 3.6.6 uses 280 data points drawn from
sampling 185 mutations across three missions.
We do not use K-fold cross validation to assess the accuracy of these models
because the training and test data are drawn from different data sets.
We test on \numexecutions data points from the data set drawn from executions of
\tool{ArduPilot} 3.6.7.
Results are in Table~\ref{table:version}.

\begin{table}[t]
\centering
\caption{Accuracy Metrics for Supervised Learning 
Trained on a Prior Version (\tool{ArduPilot} 3.6.6) and Tested on a Subsequent
Version (\tool{ArduPilot} 3.6.7).
\label{table:version}}
\begin{tabular}{r r r r | r | r}
\toprule
Mean & Mean & Mean & Mean & Num. & Num.\\
Acc. & Prec. & Rec. & F-Score & 3.6.6 & 3.6.7\\
\midrule
0.90 & 0.99 & 0.82 & 0.90 & 280 & \numexecutions \\

\bottomrule
\end{tabular}

\end{table}

By comparing the results for this experiment (Table~\ref{table:version})
with the the accuracy for the model trained and tested from the data
drawn from the same set (Table~\ref{table:supervised-total}), we find that the accuracy metrics show that this model is highly
accurate, and only slightly less accurate than a model trained and tested
on data drawn from only one version of the software.
Part of the decrease in accuracy may be attributable to the smaller
number of data points used to train this model, rather than to the different
version.

This results illustrates a promising use case for \thename in the
software development pipeline.
While developing software, a model can be trained on executions from
a prior release of the software.
While developing subsequent versions, the model can be used to assess
executions from the changed software for errors such as regressions.
This approach should apply to many situations in which the software under
development is similar to its prior release.

\subsection{RQ2: Amount of Training Data and Accuracy}

We answer RQ2: How does the volume of training data used to build 
the model affect
the accuracy of the model in detecting executions that exhibit errors?
Recall that our overall goal is to determine whether \thename
is useful in detecting software executions
that exhibit errors. This question evaluates the subgoal of
determining how much data is necessary to make useful
predictions. To answer this question, we compute the analyses
used to evaluate RQ1 in Section~\ref{sec:evaluation-rq1}
with varying amounts of data.

\begin{figure}
  \centering
  \adjustbox{max size={0.99\linewidth}{0.99\textheight}}%
            {\includegraphics{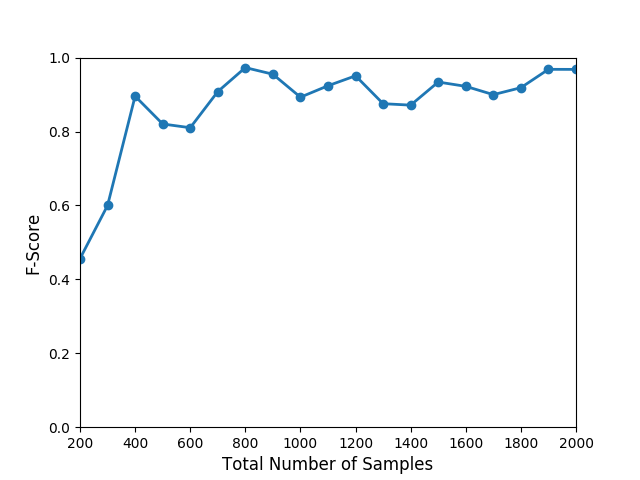}}
  \caption{\label{fig:supervised-vary-data}Supervised Learning with Varied Amounts of Data}
\end{figure}

We show results for all accuracy metrics in Table~\ref{table:supervised-vary-data} and graph
results for the F-Score metric in Figure~\ref{fig:supervised-vary-data}.
Higher is better. The X-axis
represents the total number of samples included for supervised learning.
(The samples are split between training and testing samples for K-fold
cross-validation.)
While results for 200 and 300 are middling,
F-Score quickly rises to 0.90 at 400 samples and stays above 0.80
at all sample sizes above 400.
These results show that \thename can obtain reasonable accuracy with
comparatively few data points.
Computing 400 data points would take approximately 1274.55 minutes or
less than a day.
The F-Score does not fall below 0.90 after 1500 data points.
Computing 1500 data points would take approximately 4779.55 minutes or
a little more than three days.
Such training time could become part of the workflow for a development
or testing procedure.
This is especially the case because, as demonstrated in 
Section~\ref{sec:evaluation-rq1-release}, such a model does not need to be
trained on an identical version of the software to be effective,
so a model built on an earlier version of software can continue to be used
for subsequent development.

%Note that the Y-axis starts
%at the value of 0.84 and that the lowest F-Score we observe
%is 0.8452 for the value of 50 samples. The F-Score remains
%above 0.96 for all sample sizes greater than or equal to 500.
%These data show that even with a comparatively small sample
%size of 50 samples, we get reasonably accurate data and that,
%although accuracy generally increases with additional samples,
%returns diminish after 500 samples.

\begin{table}[t]
\centering
\caption{Accuracy Metrics and Approximate Training Time for 
Varying Numbers of Data Points.
\label{table:supervised-vary-data}}
\begin{tabular}{r | r r r r | r}
\toprule
Num. & Mean & Mean & Mean & Mean & Data Gen. \\
Samples & Acc. & Prec. & Rec. & F-Score & Time (mins) \\
\midrule
%200 &  0.94 & 0.50 & 0.47 & 0.48 \\
%300 &  0.94 & 0.70 & 0.68 & 0.69 \\
%400 &  0.89 & 0.89 & 0.68 & 0.72 \\
%500 &  0.96 & 0.80 & 0.73 & 0.75 \\
%600 &  0.92 & 0.89 & 0.85 & 0.87 \\
%700 &  0.93 & 0.99 & 0.86 & 0.91 \\
%800 &  0.96 & 0.99 & 0.93 & 0.96 \\
%900 &  0.95 & 0.99 & 0.91 & 0.94 \\
%1000 &  0.93 & 0.99 & 0.87 & 0.91 \\
%1100 &  0.92 & 0.98 & 0.85 & 0.90 \\
%1200 &  0.92 & 0.99 & 0.85 & 0.90 \\
%1300 &  0.90 & 0.99 & 0.81 & 0.87 \\
%1400 &  0.93 & 0.99 & 0.86 & 0.91 \\
%1500 &  0.88 & 0.99 & 0.76 & 0.85 \\
%1600 &  0.96 & 0.99 & 0.92 & 0.95 \\
%1700 &  0.94 & 0.99 & 0.89 & 0.93 \\

%200 &  1.00 & 0.50 & 0.50 & 0.50 \\
%300 &  0.97 & 0.79 & 0.74 & 0.76 \\
%400 &  0.94 & 0.60 & 0.49 & 0.50 \\
%500 &  0.91 & 0.99 & 0.82 & 0.86 \\
%600 &  0.86 & 0.88 & 0.72 & 0.77 \\
%700 &  0.86 & 0.98 & 0.72 & 0.79 \\
%800 &  0.89 & 0.98 & 0.80 & 0.87 \\
%900 &  0.91 & 0.99 & 0.83 & 0.88 \\
%1000 &  0.95 & 0.99 & 0.91 & 0.94 \\
%1100 &  0.89 & 0.99 & 0.80 & 0.85 \\
%1200 &  0.95 & 0.99 & 0.91 & 0.94 \\
%1300 &  0.96 & 0.99 & 0.92 & 0.95 \\
%1400 &  0.92 & 0.99 & 0.84 & 0.90 \\
%1500 &  0.92 & 0.99 & 0.85 & 0.91 \\
%1600 &  0.93 & 0.99 & 0.87 & 0.92 \\
%1700 &  0.93 & 0.99 & 0.86 & 0.91 \\
%1800 &  0.94 & 0.99 & 0.89 & 0.94 \\

200 &  0.90 & 0.50 & 0.43 & 0.46 & 637.27 \\
300 &  0.89 & 0.65 & 0.59 & 0.60 & 955.91 \\
400 &  1.00 & 0.90 & 0.89 & 0.90 & 1274.55\\
500 &  0.94 & 0.89 & 0.78 & 0.82 & 1593.18 \\
600 &  0.93 & 0.89 & 0.76 & 0.81 & 1911.82 \\
700 &  0.92 & 0.98 & 0.86 & 0.91 & 2230.45 \\
800 &  0.98 & 0.99 & 0.96 & 0.97 & 2549.09 \\
900 &  0.96 & 0.99 & 0.93 & 0.96 & 2867.73 \\
1000 &  0.91 & 0.99 & 0.84 & 0.89 & 3186.36 \\
1100 &  0.94 & 0.99 & 0.89 & 0.92 & 3505.00 \\
1200 &  0.96 & 0.99 & 0.92 & 0.95 & 3823.64 \\
1300 &  0.89 & 0.98 & 0.80 & 0.88 & 4142.27 \\
1400 &  0.90 & 0.99 & 0.82 & 0.87 & 4460.91 \\
1500 &  0.94 & 0.98 & 0.90 & 0.93 & 4779.55 \\
1600 &  0.93 & 0.99 & 0.88 & 0.92 & 5098.18 \\
1700 &  0.92 & 0.99 & 0.85 & 0.90 & 5416.82 \\
1800 &  0.93 & 0.99 & 0.86 & 0.92 & 5735.45 \\
1900 &  0.97 & 0.99 & 0.95 & 0.97 & 6054.09 \\
2000 &  0.97 & 0.99 & 0.95 & 0.97 & 6372.73 \\

\bottomrule
\end{tabular}

\end{table}

\subsection{RQ3: Finding Errors Before the End of Execution}
\label{sec:evaluation-before-end}

\begin{table}[t]
\centering
\caption{Accuracy Metrics for Supervised Learning using Data at Intervals
During Execution
\label{table:supervised-before-end}}
%\begin{tabular}{r | r r r r | r | r}
\begin{tabular}{r | r r r r | r}
\toprule
%Ins. & Mean & Mean & Mean & Mean & Exec. & Num.\\
%Exec. & Acc. & Prec. & Rec. & F-Score & Time (secs) & Samples \\

Ins. & Mean & Mean & Mean & Mean & Num.\\
Exec. & Acc. & Prec. & Rec. & F-Score & Samples \\

\midrule

10000 &  0.50 & 0.00 & 0.00 & 0.00 & \numexecutions \\
20000 &  0.50 & 0.40 & 0.80 & 0.53 & \numexecutions \\
30000 &  0.57 & 0.56 & 0.36 & 0.42 & \numexecutions \\
40000 &  0.57 & 0.59 & 0.35 & 0.44 & \numexecutions \\
50000 &  0.56 & 0.63 & 0.28 & 0.37 & \numexecutions \\
60000 &  0.58 & 0.63 & 0.30 & 0.39 & \numexecutions \\
70000 &  0.56 & 0.67 & 0.32 & 0.40 & \numexecutions \\
80000 &  0.56 & 0.54 & 0.45 & 0.44 & \numexecutions \\
90000 &  0.55 & 0.52 & 0.49 & 0.47 & \numexecutions \\
100000 &  0.57 & 0.53 & 0.58 & 0.52 & \numexecutions \\
110000 &  0.52 & 0.48 & 0.21 & 0.29 & \numexecutions \\
120000 &  0.52 & 0.48 & 0.09 & 0.15 & 1708 \\
130000 &  0.71 & 0.19 & 0.12 & 0.15 & 989 \\
%\midrule
%End &  0.95 & 0.99 & 0.90 & 0.94 & \numexecutions \\

\bottomrule
\end{tabular}

\end{table}

We answer RQ3: To what extent can \thename accurately detect executions that
 exhibit errors before reaching the end of the execution?

Recall that our overall goal is to determine whether \thename is useful in 
detecting software defects.
This question evaluates the subgoal of determining how early in execution
\thename can find errors, only using data that is collected before the end of execution.
In applications in the field, it is important to find defects before
they manifest in observable crashes or other potentially-dangerous behaviors.
This could correspond to a realistic situation in which we have training data 
from known good and bad executions, from which we trained a model, and we want 
to observe a new execution and put it in to failsafe mode
if our technique classifies
the data as corresponding to bad behavior.

To answer this question, we output summaries from our DBI tool every 10,000, 
instructions, on the same corpus as used for RQ1.
We build predictive machine learning models at each interval,
for the data from all executions that reached that interval.
We chose 10,000 machine instructions as an interval after which to
output execution summaries, to trade off between overhead and
fidelity.
It corresponds to approximately eleven seconds of
execution.

We measure Accuracy, Precision, Recall, and F-Score, as described in
Section~\ref{sec:approach-supervised}.
We report the means across 10-fold cross-validation.
For each metric, higher is better and means that the technique's
determinations are more accurate.
Table~\ref{table:supervised-before-end} shows results. 

Before the end of execution, the maximum F-Score is 0.52, which does not 
reflect high predictive power.
However, the poor performance may be due to our choice to output information
at boundaries determined by the number of instructions executed.
In Section~\ref{sec:evaluation-features} and Table~\ref{table:features},
we analyze the importance of various features to the predictive power of
our models.
The count of instructions analyzed and several counts that may be related,
are highly predictive.
Our choice to keep the number of instructions analyzed constant in computing
each model before the end of execution may have inadvertently removed
variation in the most predictive signals.
This suggests that a similar approach may have a better chance of success
by looking at more natural boundaries based on properties of the software.

\subsection{RQ4: Overhead}
\label{sec:evaluation-overhead}
We answer RQ4: How much overhead is introduced by DBI?

Recall that our overall goal is to determine whether \thename
is useful in detecting software executions
that exhibit errors.
This question evaluates the subgoal of
determining the amount of overhead involved in making these predictions. 
Knowing the amount of overhead
aids in determining whether the technique will be useful in
various real-world situations.

To answer this question, we measure the overhead of \dbitool tool on executions of \tool{ArduPilot}.
To compute the overhead, we ran \tool{ArduPilot} with a representative
input mission and mutation several times, with each of three variations
on
instrumentation: No instrumentation, \tool{Valgrind}'s example
\tool{Lackey} tool,
and our custom tool \dbitool based on the \tool{Valgrind} platform.\footnote{We do
not evaluate \tool{Valgrind}'s better-known \tool{Memcheck} tool because of a technical
limitation.}
We timed the seconds elapsed for each execution.
We measured a 24\% increase in execution time over no instrumentation
when we instrumented \tool{ArduPilot} with \tool{Lackey}.
However, to the resolution we measured, \tool{ArduPilot} running with our
custom instrumentation tool did not take longer to run than when running
without instrumentation.
Under these circumstances, the overhead of our custom tool is negligible.

%In addition, we report the amount of time it takes to build the models

% On the same ls command:
% uninstrumented: 0.002
% lackey: 0.807
% memcheck: 0.766
% debgrind: 0.371

% On ArduPilot
% None lackey memcheck debgrind

% None
% 2fd69cdb
% 10:01:11,582 - 10:08:12,243 : 7:01 = 421
% 10:21:48,808 - 10:28:34,411 : 6:46 = 406
% 10:42:10,416 - 10:48:55,811 : 6:46 = 406  avg: 411

%% lackey
% 11:02:32,593 - 11:11:02,718 : 8:30 = 510
% 11:28:08,132 - 11:36:38,953 : 8:31 = 511
% 11:53:44,659 - 12:02:14,257 : 8:29 = 509 avg: 510 1.2x

% Memcheck

% Debgrind
% 14:28:05,470 - 14:34:54,766 : 6:49 = 408
% 14:39:41,504 - 14:46:29,762 : 6:48 = 407
% 14:46:33,730 - 14:53:22,296 : 6:48 = 407 avg: 407

Recall that running a similar test on a standard Linux utility
produced overheads of: \tool{Lackey}: 404x; \tool{Memcheck}: 383x; and
our custom tool: 186x.
These numbers show that dynamic binary instrumentation can incur far less
overhead, as measured by time, in robotics systems than in standard
CPU- and I/O-bound programs.
%\todo{collect these numbers across more executions and automate their
%collection}

%\todo{source instrumentation experiment explanation and results}

\subsection{RQ5: Effects of Delays on Robot Behavior}
\label{sec:evaluation-delays}
We answer RQ5: What effects do delays have on observable robot performance?

Recall that our overall goal is to determine whether \thename is useful
in detecting software executions that exhibit errors.
This question evaluates the subgoal of evaluating the effects of overhead
on the behavior of the robot.
When overhead can be absorbed with little observable effect,
\thename is most useful.
To do so, we evaluate the following sub-questions:
\begin{description}
\item[\textbf{RQ5a:}] To what extent do the presence of timing delays in robot systems have an effect on observable behavior as defined by a set of performance metrics?
\item[\textbf{RQ5b:}] Under what circumstances do timing delays lead to system crashes?
\end{description}

As shown above, \acfp{ARS} are amenable to detection of faults by the use of
low-level program monitoring.
One primary concern about using these types of
monitoring techniques is that the techniques can cause high overhead.
\acp{CPS} such as \acp{ARS} can be sensitive to overhead that interferes with
the timing of events --- a missed deadline
or a sequence of messages received in an unexpected order can cause the system to fail.
However, at the same time, these systems are particularly prone to variability in
operating conditions because of their interaction with the real world and the
unpredictable conditions therein.
There are many situations in which the architectures of \acp{CPS} can absorb timing
delays, when they
take place during times when the system would otherwise be spent waiting for physical
events or communication from other parts of the system.

This research question evaluates the nature and extent of the timing delays
that \acp{ARS} can absorb.
To do so, we conduct a series of experiments in simulation
to gain a more precise understanding of the amount
and nature of delays that these systems can absorb.
The nominal executions examine the behavior of an unmodified simulated \ac{ARS}
while the executions with artificial delays examine the behavior of the same systems
when message passing is delayed for various topics.

\paragraph{Timing Delay Metrics}
\label{sec:timing-methodology-metrics}
To evaluate the effects of timing delays on observable robotics behavior, we use the following metrics, based in Euclidean distance, completeness, and timeliness.

\begin{itemize}

%\item The Euclidean distance between the final position of the robot in the representative nominal and each deformed execution
\item The Euclidean distance between the final position of the robot and the final waypoint or home point.
%\item Given aligned time series between the representative nominal execution and each deformed execution, the greatest and the average Euclidean difference between each position on the path
\item The sum of closest Euclidean distances on the trajectory from each waypoint.
\item The mean of the closest Euclidean distances on the trajectory from each waypoint.
%\item The greatest closest distance from each waypoint.

\item Whether the execution navigates to each waypoint and returns home.
\item The amount of time before completion of the execution (either successful or unsuccessful).

\end{itemize}

\subsubsection{Effects on Observable Behavior}
\label{sec:timing-methodology-RQ-observable}

\textbf{RQ5a:} To what extent do the presence of timing delays in robot systems have an effect on observable behavior as defined by a set of performance metrics?
To evaluate RQ5a, we look at the metrics enumerated above.

The clearest and most obvious effects on observable execution are crashes, both software crashes and crashes in physical space.
We evaluate these deviations separately in RQ5b (Section~\ref{sec:timing-methodology-RQ-crash}).

\newcommand\multiVI[1]{\multicolumn{6}{c}{#1}}

\paragraph{\tool{Husky}}

Table~\ref{table:husky-euclidean-distance} shows, for the nominal and 
artificially-deformed 
\tool{Husky} executions, how much their Euclidean distance deviates from the 
waypoints the robot had been instructed to visit.
%and the trajectory of the representative nominal execution.
Data for each mission is listed on its own line.
For the purposes of this chart, we look at the trajectories of all experimental
runs that reach all of the waypoints for a given mission. 
We take the robot's minimum distance from each waypoint for each of these experimental 
executions.
We then take the mean, over all of these experimental executions, of the minimum
distance for each waypoint.
%\todo{report std. deviation too?}
The same information is provided for the nominal executions for comparison.
Note that the experimental executions include varying amounts of delay and
delays on different \tool{rostopics}.
We will explore the effects of different delay amounts and delays on different topics
in Table~\ref{table:husky-delay-variable}.

Note that in Table~\ref{table:husky-euclidean-distance}, the mean closest distance
to the waypoint is always smaller in the nominal group (which is taken from
unmodified executions) than in the experimental group (which is taken from
executions with delays).
This shows that \tool{Husky}'s operation in simulation is sensitive to artificial
delays. 
However, the mean closest distance for each waypoint in the experimental data
is almost always within a reasonable tolerance --- the robot gets reasonably close
to its target.
Furthermore, the missions and waypoints for which the distances are higher for
the experimental runs are also the same missions and waypoints for which the distances
are higher for the nominal runs.
For example, for M1, the smallest mean distance from the waypoint is for W1 while the
largest is for W5, in both the nominal and the experimental data.
This implies that the delays enhance the existing effects of which waypoints \tool{Husky} finds easier or difficult to navigate to.
While the closest distance from the destination waypoint usually increases
over subsequent waypoints, that is not always the case.

\begin{table*}[t]
\centering
\caption{Husky: Mean Minimum Euclidean Distance from Waypoints in Meters for Nominal Executions and Experimental Executions 
\label{table:husky-euclidean-distance}}
\begin{tabular}{r | r r r r r r | r r r}%| r r}
\toprule
Mission & \multiVI{Distance from Waypoint} & WP & WP \\%& Nominal & Nominal \\
        & W1 & W2 & W3 & W4 & W5 & Final & Total & Mean \\%& Final & Mean \\
\midrule
%Nominal \\
%M1 & 0.23 & 0.66 & 1.65 & 1.47 & 2.01 & 1.10 & 7.13 & 1.19 \\
%M2 & 0.08 & 0.18 & 0.25 & 0.15 & 0.30 & 0.41 & 1.37 & 0.23 \\
%M3 & 0.22 & 0.38 & 0.54 & 0.58 & 2.85 & 1.45 & 6.02 & 1.00 \\
%M4 & 0.52 & 1.91 & 1.78 & 2.28 & 1.02 & 1.92 & 9.44 & 1.57 \\
%M5 & 0.39 & 1.01 & 1.12 & 0.57 & 2.09 & 2.67 & 7.86 & 1.31 \\
%M6 & 0.25 & 0.31 & 0.70 & 0.27 & 0.76 & 1.44 & 3.71 & 0.62 \\
%M7 & 0.31 & 0.27 & 0.28 & 0.35 & 0.42 & 0.55 & 2.18 & 0.36 \\
%M8 & 0.28 & 0.45 & 0.23 & 0.98 & 0.68 & 0.95 & 3.57 & 0.60 \\
%M9 & 0.80 & 0.53 & 0.75 & 0.59 & 0.83 & 3.43 & 6.93 & 1.16 \\
%M10 & 0.07 & 0.07 & 0.08 & 0.66 & 1.09 & 1.91 & 3.89 & 0.65 \\
%\midrule
%Experimental \\
%M1 & 0.83 & 1.13 & 3.14 & 2.69 & 3.37 & 1.79 & 12.96 & 2.16 \\
%M2 & 0.72 & 0.45 & 1.37 & 0.83 & 1.20 & 0.93 & 5.50 & 0.92 \\
%M3 & 1.40 & 1.84 & 1.84 & 1.93 & 4.53 & 3.22 & 14.76 & 2.46 \\
%M4 & 1.33 & 2.91 & 3.23 & 3.14 & 1.97 & 2.28 & 14.85 & 2.47 \\
%M5 & 1.69 & 1.95 & 2.39 & 1.60 & 3.39 & 3.44 & 14.46 & 2.41 \\
%M6 & 0.99 & 1.29 & 2.56 & 0.75 & 2.81 & 1.99 & 10.39 & 1.73 \\
%M7 & 1.44 & 0.77 & 1.04 & 1.55 & 1.98 & 1.52 & 8.30 & 1.38 \\
%M8 & 1.21 & 2.86 & 1.92 & 3.94 & 2.74 & 1.87 & 14.54 & 2.42 \\
%M9 & 1.06 & 0.66 & 1.01 & 0.83 & 0.99 & 2.88 & 7.43 & 1.24 \\
%M10 & 2.03 & 1.37 & 0.28 & 1.93 & 2.33 & 3.82 & 11.77 & 1.96 \\
Nominal \\
M1 & 0.23 & 0.66 & 1.65 & 1.47 & 2.01 & 1.10 & 7.13 & 1.19 \\
M2 & 0.08 & 0.18 & 0.25 & 0.15 & 0.30 & 0.41 & 1.37 & 0.23 \\
M3 & 0.22 & 0.38 & 0.54 & 0.58 & 2.85 & 1.45 & 6.02 & 1.00 \\
M4 & 0.52 & 1.91 & 1.78 & 2.28 & 1.02 & 1.92 & 9.44 & 1.57 \\
M5 & 0.39 & 1.01 & 1.12 & 0.57 & 2.09 & 2.67 & 7.86 & 1.31 \\
M6 & 0.25 & 0.31 & 0.70 & 0.27 & 0.76 & 1.44 & 3.71 & 0.62 \\
M7 & 0.31 & 0.27 & 0.28 & 0.35 & 0.42 & 0.55 & 2.18 & 0.36 \\
M8 & 0.28 & 0.45 & 0.23 & 0.98 & 0.68 & 0.95 & 3.57 & 0.60 \\
M9 & 0.80 & 0.53 & 0.75 & 0.59 & 0.83 & 3.43 & 6.93 & 1.16 \\
M10 & 0.07 & 0.07 & 0.08 & 0.66 & 1.09 & 1.91 & 3.89 & 0.65 \\
\midrule
Experimental \\
M1 & 1.00 & 1.20 & 3.56 & 2.92 & 3.61 & 1.89 & 14.19 & 2.37 \\
M2 & 0.89 & 0.52 & 1.67 & 1.00 & 1.40 & 1.12 & 6.60 & 1.10 \\
M3 & 1.85 & 2.38 & 2.25 & 2.34 & 5.04 & 3.48 & 17.34 & 2.89 \\
M4 & 1.42 & 2.83 & 3.26 & 3.06 & 2.03 & 2.34 & 14.94 & 2.49 \\
M5 & 1.86 & 2.01 & 2.48 & 1.72 & 3.56 & 3.73 & 15.36 & 2.56 \\
M6 & 1.03 & 1.34 & 2.67 & 0.79 & 2.96 & 2.05 & 10.83 & 1.80 \\
M7 & 1.61 & 0.86 & 1.15 & 1.73 & 2.21 & 1.69 & 9.24 & 1.54 \\
M8 & 1.21 & 2.85 & 1.87 & 3.85 & 2.70 & 1.93 & 14.42 & 2.40 \\
M9 & 1.41 & 0.87 & 1.32 & 1.07 & 1.25 & 3.10 & 9.02 & 1.50 \\
M10 & 2.57 & 1.73 & 0.34 & 2.39 & 2.68 & 4.12 & 13.82 & 2.30 \\

\bottomrule
\end{tabular}

\end{table*}
%\todo{MAKE THIS A BOX PLOT INSTEAD? SOMETHING TO BETTER MAKE THE COMPARISONS?}
%M1 HUSKY_pose_array_5_156109eab4cc4c1f9432690d0b6e6ca9.yaml
%M2 HUSKY_pose_array_5_c8629fc042474a9db240282a52448964.yaml
%M3 HUSKY_pose_array_5_5f31cf0a67be40a4b9adcb331f43361c.yaml
%M4 HUSKY_pose_array_5_350028c9199b4d15b236d85c10ff8b9e.yaml
%M5 HUSKY_pose_array_5_2ce0b4fa2a4249b4b5057913b8c3d9ec.yaml
%M6 HUSKY_pose_array_5_c787950abab14110836ae170213406d8.yaml
%M7 HUSKY_pose_array_5_865a8b7817474e51861522bd435faff1.yaml
%M8 HUSKY_pose_array_5_722ba706e1f34beca0bcaf193d7e0c4f.yaml
%M9 HUSKY_pose_array_5_cdc8edbd1c2b4ce197a211ee64e7cbe4.yaml
%M10 HUSKY_pose_array_5_de68abeac0924afc938d17e5acd1b825.yaml
%M11 HUSKY_pose_array_5_a4d27db27974459a97a696b7d98cb403.yaml

\begin{table*}[t]
\centering
\caption{Husky: Mean Minimum Euclidean Distance (in Meters) from Waypoints for Varied Delays
on Mission 1, Topic \tool{/husky\_velocity\_controller/odom}
\label{table:husky-delay-variable}}
\begin{tabular}{l | r r r r r r | r r }%| r r}
\toprule
Delay & \multiVI{Distance from Waypoint} & WP & WP \\%& Nominal & Nominal \\
(s)      & W1 & W2 & W3 & W4 & W5 & Final & Total & Mean \\%& Final & Mean \\
\midrule

Mean \\
0.0 & 0.23 & 0.66 & 1.65 & 1.47 & 2.01 & 1.10 & 7.13 & 1.19 \\
0.00390625 & 4.45 & 3.59 & 10.63 & 8.84 & 8.70 & 4.50 & 40.71 & 6.79 \\
0.015625 & 4.44 & 3.58 & 10.61 & 8.83 & 8.75 & 4.51 & 40.72 & 6.79 \\
0.0625 & 4.42 & 3.57 & 10.55 & 8.81 & 8.66 & 4.47 & 40.48 & 6.75 \\
0.25 & 4.43 & 3.59 & 10.57 & 8.81 & 8.65 & 4.44 & 40.49 & 6.75 \\
1.0 & 4.45 & 3.60 & 10.65 & 8.87 & 8.73 & 4.47 & 40.77 & 6.80 \\
\midrule
Standard Deviation \\
0.0 & 0.41 & 1.13 & 2.62 & 2.89 & 3.49 & 1.80 & 12.35 & 2.06 \\
0.00390625 & 0.86 & 0.64 & 1.87 & 1.65 & 1.56 & 0.81 & 7.39 & 1.23 \\
0.015625 & 0.90 & 0.67 & 1.88 & 1.61 & 1.39 & 0.77 & 7.21 & 1.20 \\
0.0625 & 0.93 & 0.71 & 2.08 & 1.70 & 1.67 & 0.81 & 7.89 & 1.32 \\
0.25 & 0.91 & 0.66 & 2.02 & 1.73 & 1.68 & 0.85 & 7.84 & 1.31 \\
1.0 & 0.86 & 0.63 & 1.79 & 1.54 & 1.47 & 0.80 & 7.09 & 1.18 \\

\bottomrule
\end{tabular}

\end{table*}

\subsubsection{When Delays Cause Software Crashes}
\label{sec:timing-methodology-RQ-crash}

\textbf{RQ5b:} Under what circumstances do timing delays lead to system crashes?

To evaluate RQ5b, we look at several indicators of software crashes that can be observed from experiments.
It is interesting to find out when timing causes a system crash because system crashes have different practical implications for recovery techniques than other failures, such as incorrect trajectories or delays. System crashes can lead to, for example, losing contact with the system or damage to the hardware. Under some circumstances, a system that has crashed without hardware damage can simply be restarted.
It is important to separate system crashes from other successful executions so that we can exclude any trajectories and timing data that are invalid because of system crashes.

We establish a baseline of software crashes that occur in the nominal data set. Robotics systems are often nondeterministic and difficult to simulate and,
therefore, even nominal executions can experience software crashes.
We compare the rate of software crashes in nominal executions against the rate of software crashes under the experimental conditions.

The presence of a core dump file --- such as would be produced when a segmentation fault occurs --- indicates a system crash
The absence of logs that would have normally been produced during a proper execution indicates a system crash.
If the test harness exits abnormally, we classify that execution as a system crash.

\paragraph{Results}

Table~\ref{table:husky-crashes} shows the percent of \tool{Husky} executions
that either crash or do not reach the end goal (within a tolerance of one meter).
For the purposes of these results, we group all executions that do not reach the final
goal within the established tolerance for any reason.
Here, failure to reach the end goal within the established tolerance is a proxy for
the execution having crashed.
It is based on the assumption that a system crash will occur before the end of the 
designated mission and prevent the robot from reaching its goal.
There is also an assumption that, if the robot has gone so badly wrong that it does not 
reach its goal within the established tolerance, it is functionally equivalent to crashing.

\begin{table*}[t]
\centering
\caption{Husky: Percent of Executions That Crash or Do Not Reach all Waypoints on Mission 1, With a Tolerance of One meter
%TODO: change this to not reaching end goal. or better yet, show both.
\label{table:husky-crashes}}
\begin{tabular}{l | r r r r r r  r r }%| r r}
\toprule
Topic & \multiVI{Percent Failure by Delay Amount}  \\%& Nominal & Nominal \\
      & 0.0 & 0.00390625 & 0.015625 & 0.0625 & 0.25 & 1.00\\%& Final & Mean \\
\midrule

/gazebo/link\_states & 60.08 & 2.50 & 0.83 & 2.50 & 5.00 & 2.50 & \\
/husky\_velocity\_controller/cmd\_vel & 60.08 & 3.33 & 2.50 & 1.67 & 4.17 & 4.17 & \\
/husky\_velocity\_controller/odom & 60.08 & 2.50 & 0.83 & 4.17 & 0.00 & 4.17 & \\
/imu/data & 60.08 & 0.00 & 2.50 & 3.33 & 3.33 & 2.50 & \\
/imu/data/bias & 60.08 & 1.67 & 1.67 & 3.33 & 1.67 & 0.83 & \\
/navsat/fix & 60.08 & 3.33 & 1.67 & 3.33 & 3.33 & 0.00 & \\

\bottomrule
\end{tabular}
\end{table*}

\newcommand\multiX[1]{\multicolumn{10}{c}{#1}}

\begin{table*}
\centering
\caption{Husky: Mean Time Taken (seconds) for Executions That Reach (P) and Do Not Reach (F) the Final Waypoint on Mission 1, With a Tolerance of One Meter
\label{table:husky-time}}
\begin{tabular}{l | r r | r r | r r | r r | r r | r r r r r r}
\toprule
Topic & \multiX{Delays (Seconds)}  \\
(abbreviated)  & \multiII{0.00390625} & \multiII{0.015625} & \multiII{0.0625} & \multiII{0.25} & \multiII{1.0}\\%& Final & Mean \\
      & P & F & P & F & P & F & P & F & P & F  \\
\midrule

/gazebo/link\_states &  27.45 &  33.81 &  28.14 &  33.62 &  26.97 &  33.84 &  27.91 &  33.67 &  27.13 &  33.69  \\
/husky.../cmd\_vel &  27.03 &  33.77 &  27.82 &  33.72 &  27.74 &  33.58 &  27.39 &  33.76 &  27.35 &  33.71  \\
/husky.../odom &  27.36 &  33.77 &  26.49 &  33.59 &  27.42 &  33.65 & n/a &  33.61 &  27.23 &  33.88  \\
/imu/data & n/a &  33.67 &  27.59 &  33.70 &  27.60 &  33.69 &  27.32 &  33.85 &  27.06 &  33.70  \\
/imu/data/bias &  28.07 &  33.86 &  27.58 &  33.66 &  27.69 &  33.76 &  27.63 &  33.68 &  28.44 &  33.58  \\
/navsat/fix &  27.11 &  33.68 &  28.21 &  33.63 &  28.01 &  33.75 &  27.67 &  33.80 & n/a &  33.70  \\

\bottomrule

\end{tabular}

\end{table*}

\subsection{RQ6: Feature Importance and Overhead}
\label{sec:evaluation-features}
We answer RQ6: To what extent does feature reduction affect overhead and
 accuracy?
 
Recall that our overall goal is to determine whether \thename is useful
in detecting software executions that exhibit errors.
This question evaluates the subgoals of (1) determining the low-level
signals that best contribute to making accurate determinations 
and (2) determining if limiting data collection to those most informative
signals would reduce overhead in data collection while maintaining accuracy.

To determine which low-level signals have the most predictive power,
we examine the properties of the classifiers we build to answer RQ1.
The trained decision tree classifiers from Scikit Learn have a property --
\texttt{feature\_importances} -- which, for each feature in the input
data, returns a floating point number corresponding to that feature's importance
in the algorithm building the classifier.
Because we use 10-fold cross validation in building and validating the
model, we have a set of ten lists of feature importances.
Each list contains 19 or 20 features with importance zero, which means
they were not used in the classifier.

We summarize data on the features with a non-zero importance in computing
the models for any of the ten folds in Table~\ref{table:features}.
StoreCount, WrTmpCount, and ExitCount are tallies of the number of
instructions executed that \tool{Valgrind} categorizes as
\texttt{Ist\_Store}, \texttt{Ist\_WrTmp}, and \texttt{Ist\_Exit},
respectively.
InsCount is a tally of the total number of instructions executed.
SBExit and SBEnter are tallies of how many times \tool{Valgrind} records 
entry into and exit from a superblock, which \tool{Valgrind} defines
to be, "a single entry, multiple exit, linear chunk of 
code."\footnote{\url{http://valgrind.org/docs/manual/lk-manual.html}}
MaxInsAddr is the highest address of a machine instruction executed.
InsAddrDiff is the difference between the maximum instruction address
and the lowest address of a machine instruction executed.

To compute accuracy, we re-run model creation for the question we ask
in RQ1 using the original data but restricting the algorithm to only
use the same five signals.
Results are in Table~\ref{table:feature-accuracy}.
As one can see by comparing this table, to 
Table~\ref{table:supervised-total}, the model
built with only the five features is nearly as accurate as the model
built using all 26.
This result is not surprising given that the decision tree algorithm 
considered these five features most important when building its models.
The result suggests that, once one determines which features are most
useful for a particular category of program, instrumentation can be limited
to collecting data for those features without significant loss of accuracy.

To compute overhead, we create a new custom \tool{Valgrind} tool
that only collects the five signals that are identified as most important.
We time the execution of this new tool across eight mutations and
three missions and compare the timing on these same mutations and
missions for our original customized tool.
We find that the tool with the reduced feature set does not consistently
save time over the tool that collects all 26 features.
In fact, the reduced feature tool
often takes longer than running the full feature tool.
The reduced feature tool takes longer in 7 cases, is about the same
as the original tool in 7 cases, and saves time in the remaining 10 cases.
On average the reduced feature tool took 0.8 seconds longer to run,
with it taking 10 seconds longer in the worst case.
This result is also not surprising given that all of the
features had been collected within the context of an already optimized tool
and the same basic instrumentation was necessary to collect the five features
as the 26.

\begin{table}[t]
\centering
\caption{Feature Importance across Ten Folds
\label{table:features}}
\begin{tabular}{l | r | r}
\toprule
Signal & Number of Folds & Mean Importance \\
\midrule
StoreCount & 10 & 0.4528 \\
WrTmpCount & 10 & 0.2721 \\
InsCount & 10 & 0.2178 \\
SBExit & 10 & 0.0374 \\
ExitCount & 10 & 0.0111 \\
SBEnter & 9 & 0.0093 \\
InsAddrDiff & 7 & 0.0004 \\
MaxInsAddr & 3 & 0.0002 \\

\bottomrule
\end{tabular}

\end{table}

\begin{table}[t]
\centering
\caption{Accuracy Metrics for Supervised Learning 
End of Execution Using only the Five Most Predictive Signals.
\label{table:feature-accuracy}}
\begin{tabular}{r r r r | r}
\toprule
Mean & Mean & Mean & Mean & Num.\\
Acc. & Prec. & Rec. & F-Score & Samples \\
\midrule
0.93 & 0.99 & 0.87 & 0.92 & \numexecutions \\

\bottomrule
\end{tabular}

\end{table}

%\begin{table}[t]
%\centering
%\caption{Timing for Feature-Reduced DBI\label{table:feature-time}}
%\begin{tabular}{r r r r r}
%\toprule
%Mission & Mutation & Full Tool & Reduced Tool & Difference\\
%ID & ID & Time (secs) & Time (secs)\\
%914b17b8 & 1e9c25e4 & 409 & 417 & -8 \\
%914b17b8 & c10e6f9c & 128 & 129 & -1 \\
%914b17b8 & b7bb2a21 & 127 & 134 & -7 \\
%914b17b8 & b4035820 & 127 & 127 & 0 \\
%914b17b8 & 19743fd8 & 128 & 127 & 1 \\
%914b17b8 & ad916135 & 128 & 138 & -10 \\
%914b17b8 & 523f736b & 127 & 130 & -3 \\
%914b17b8 & f958cc21 & 127 & 128 & -1 \\
%71bfcb56 & 1e9c25e4 & 403 & 401 & 2 \\
%71bfcb56 & c10e6f9c & 448 & 448 & 0 \\
%71bfcb56 & b7bb2a21 & 448 & 447 & 1 \\
%71bfcb56 & b4035820 & 448 & 447 & 1 \\
%71bfcb56 & 19743fd8 & 449 & 448 & 1 \\
%71bfcb56 & ad916135 & 448 & 447 & 1 \\
%71bfcb56 & 523f736b & 448 & 447 & 1 \\
%71bfcb56 & f958cc21 & 449 & 448 & 1 \\
%f18750e4 & 1e9c25e4 & 403 & 402 & 1 \\
%f18750e4 & c10e6f9c & 166 & 165 & 1 \\
%f18750e4 & b7bb2a21 & 165 & 165 & 0 \\
%f18750e4 & b4035820 & 165 & 165 & 0 \\
%f18750e4 & 19743fd8 & 165 & 165 & 0 \\
%f18750e4 & ad916135 & 165 & 165 & 0 \\
%f18750e4 & 523f736b & 165 & 165 & 0 \\
%f18750e4 & f958cc21 & 165 & 166 & -1 \\
%
%
%\bottomrule
%\end{tabular}
%\end{table}
%

\section{Discussion and Threats to Validity}

This section discusses future directions and implications of the experiments
presented here, along with threats to validity.

\subsection{Timing}

The experiments on timing delays yield interesting possible future directions.

\subsubsection{Future Directions}

There are several questions that arise directly from the work presented here.

\paragraph{Violations of Other Desired Properties}

The work presented here looks at the extent to which artificial timing delays deform execution
in robotics programs in simulation by looking at whether the software crashes
and physically-observable properties, such as how far the robot is from the expected
position in physical space and how long the robot takes to reach waypoints.
However, there are other desired properties in robotics execution.
For example, there are safety properties that robots should maintain during
execution, such as that they should not crash into an obstacle or that they
should not violate speed limits.
In addition, robots should maintain liveness --- they should not time out.
It would be interesting to investigate the extent to which timing delays cause these properties to be violated.

\paragraph{Error Handling and Desired Corner Case Behavior}

It would be further interesting to investigate to what extent timing delays cause
robotics systems to enter into error-handling behavior.
For example, many systems are designed with \emph{fail safe} behavior, in which
the robot is designed to shut down in a non-damaging state when the system encounters
an unrecoverable error.
Error handling for less severe faults may cause the robot to execute a recovery
behavior, such as clearing its position and using its sensors to attempt to
identify where it is with respect to its environment.
Such a recovery behavior can occur even in nominal execution and is a normal part
of providing resiliency and accounting for nondeterminism in normal robotics executions.
However, timing delays may cause these behaviors to be more frequent (because
the timing delays may cause errors).

\paragraph{Examination of Variation in Nominal Behavior}
\acp{ARS} are noisy.
There is considerable variation in their nominal behavior, especially when a
perception system, an autopilot system, and obstacles are involved. 
This leads to considerable variation in paths taken by an unmodified system.
The unmodified system sometimes fails to reach all waypoints or simply gets stuck.
Additional work should examine the expected amount of variation in unmodified systems
and the causes of that variation.

\paragraph{Varied Amounts of Timing Delays}
The strategy for inserting timing delays in these experiments is relatively
simple --- a constant delay amount added to every message in the \tool{\ac{ROS}}
experiments and a constant delay amount added before probabilistically selected
return statements in the \tool{ArduPilot} experiments.
More targeted delay injections may reveal more precisely the circumstances under
which overhead is absorbed versus produces observable behavior deviations.

\subsubsection{Discussion of Timing Amounts}
\paragraph{Amount of Timing Delays as Compared to Expected Event Frequencies}
When systems expect events to occur at a given frequency, such as when there is a
control loop, a timing delay greater than the given frequency will almost certainly
cause unintended behavior.
This is reflected in these experiments, as the timing delays were chosen without
regard to the various control loop and other expected frequencies in the
underlying systems.
A portion of the delays are smaller than the various expected frequencies, 
while a portion of them are larger.
Smaller delays, when incurred multiple times in the same program region, can
translate into larger delays.
There is, however, redundancy and fault tolerance built into many \acp{ARS}.
A delay greater than an expected event frequency may appear to be absorbed
when the redundancy behaviors mask it.

\paragraph{Amount of Timing Delays as Compared to Instrumentation Delays}
These timing delays are intended to mimic delays caused by instrumentation
and monitoring.
While the timing delays inserted are not chosen by exact measurement to make them
congruent with monitoring delays, they mimic those delays in other ways.
The timing delays caused by monitoring are very small and occur very frequently ---
at every machine instruction.
The timing delays inserted in these experiments are generally larger, but they
occur less frequently.
They are intended as a rough approximation to explain the principle behind
why monitoring delays can be absorbed.
Extensions of these experiments could be used to designate practical tolerance levels for
monitoring and translate those tolerance levels into actual monitoring tools that
work within those boundaries.

\subsection{Taxonomy of Faults}
We classify instances of incorrect software behavior with reference to earlier
taxonomies.
While, colloquially, the words, ``bug,'' ``fault,'' ``error,''
``failure,'' and, ``off-nominal,''  refer to any software behavior that is
unusual and unintended, other work has broken down the nature of unintended
software behavior into a more precise taxonomy and dealt with the
classification of these types of behaviors~\cite{AvizienisTaxonomy2004, CotroneoTriggers2013, SahooServer2010, LiChanged2006, ShanMistakes2008, HenningssonFaultClassification2004, AsadollahConcurrency2015, ElEmamRepeatability1998, GrottkeSpace2010, SteinbauerRobocup2013}.
As in Avizienis et al., a service failure occurs when when program's behavior
-- or delivered service -- differs from the correct~\cite{AvizienisTaxonomy2004}
program behavior.
Note that with complex robotics and autonomous systems, it is not always
easy or, in fact, possible to determine the the exact constraints
of the correct program behavior~\cite{FraadeBlanarMeasuring2018,
KoopmanFramework2018}.
These service failures are observed in their external manifestations as
errors.
The underlying cause of an error is called a
fault~\cite{AvizienisTaxonomy2004}.
In this sense, the experiments conducted for this paper analyze service
failures that occur in executing software.
These failures originate as faults in the underlying source code.
Avizienis et al. further establish eight fault dimensions based on features such as
objective and persistence.
These fault dimensions are useful for classifying the nature of faults.
We do not make claims about whether our technique is better at identifying
faults depending on their classification.

\subsection{Threats to Validity}
\label{sec:threats}

Supervised machine learning requires that a portion of the data have labels,
so that it can be used for training.
This requires an oracle for at least a portion of the test inputs, limiting the potential
generalizability of the technique.  However, we envision its applicability in a
developer-support setting in contexts where programs are developed over long
periods of time, such that certain regression tests have already been created
with oracle outputs and can be used to support the creation of new test inputs
via automated input generation. 
Our approach also generally assumes that unusual program behavior 
corresponds to unintended behavior, which may not always be the case,
such as in corner cases or error handling code. 
Our technique may work best as part
of a suite of anomaly detection approaches, or including human oversight.

Technically, all versions of each program must be compiled with the same compiler
and compiler flags, within the same environment.
A difference in compilation across versions of the same program will cause 
differences in the signals that can cause the machine learning techniques to 
pick up on the differences in compilation rather than the differences in 
intended versus unintended behavior.
Similarly, the work assumes that the varied executions of the programs occur on
the same machines, in what is relatively the same environment.
Future work may allow signal normalization to reduce 
sensitivity to different compilation, machines, or environment.

As with any machine learning work, there is always a threat of overfitting:
that the classifiers learn specific traits from the data that do not correspond
to what we intend them to learn.  
This threat is amplified by the underlying unbalanced data set:
there are many more executions that do not exhibit errors than executions that
do.
We take several steps to mitigate this threat,
including balancing the data sets, and
conducting K-fold cross validation, indicating that our models generalize.
For certain errors, there may be simpler ways to identify anomalous behavior
than by using our models, such as timing information.  
However, our technique is more general, in that it does not rely on bug-class-specific
characteristics, and we demonstrate it on a variety of executions that exhibit
errors, to substantiate generalizability. 

There is a risk that our results may not generalize beyond the systems studied.
The \tool{ArduPilot} system has many interesting properties for the purposes of
our study. It is a mature open source project and is widely used by both
professional and hobbyist roboticists.
However, it operates on a relatively-simple control loop
design; systems with more complex architectures may have bugs that are less
amenable to being replicated and tested in simulation and detected
using these techniques.
Addressing this risk motivates future work isolating
portions of systems for observation.

Along the same lines,
these experiments are limited to the input data provided, which includes relatively
simple simulated environments and missions.
More complex behaviors may not have been tested.
Furthermore, the timing delay experiments did not test effects other than deviations in the observed three-dimensional position of the robot. 
It is possible that delays can affect other properties.
However, this threat is mitigated by the idea that any major failures in robot
execution are likely to affect three-dimensional position.

We conduct all experiments in simulation.
While it is possible to gain many insights about robotics in simulation~\cite{TimperleyCrashing2018, SotiropoulosNavigationBugs2017},
simulation may not accurately reflect some aspects of real hardware,
such as the influence of overhead on timing.
For example, real robotics hardware often has distributed computing resources
which may not be accurately reflected in the centralized computing power available in simulation.
A component with less computing power may encounter bottlenecks that are not
seen in simulation.
Simulation also has imperfect fidelity to real world situations~\cite{AfzalSimulation2020}.
However, this threat is mitigated by the fact that much of the monitoring
and bug detection can also take place in simulation.

\section{Related Work}
\label{sec:related-work}

% SchulteAssembly2010, SchulteRouter2015, SchulteExact2018, SchulteEmbedded2013, SchulteEnergy2014, ForrestSelf1996, Cylanceoptics2017, WildeReconnaissance1995, EisenbergDynamic2005

\paragraph{Dynamic analysis}
Several other dynamic analysis techniques do not require source code.
Eisenberg et al.~\cite{EisenbergDynamic2005} introduce using dynamic analysis
to trace program functionality to its location in binary or source code.
However, as with many dynamic analysis tools, the implementation is limited
to Java.
Clearview extends invariant inference and violation work to Windows x86
binaries, without the need for source code or  
debugging information~\cite{PerkinsClearview2009}.  Clearview focuses on
particular types of attacks, and is primarily designed to repair errors.  We do
not repair errors, but our technique is generic to a variety of
bug or error types.  The techniques are therefore best viewed as complementary
to one another.
\emph{Observation-based testing} techniques~\cite{DickinsonFailure2001} use
instrumentation to support dynamic analysis on existing sets of program
inputs~\cite{LeonVisualization2000}. 

% AliabadiArtinali2017, NguyenNumericalInvariants2017,  LorenzoliBehavioral2008, BeschastnikhTemporal2011, RatcliffInvariants2011, 
Dynamic invariant detection techniques automatically identify properties that
hold true over all correct executions of a program, and identify bugs via
violation of those invariants.  Well-known techniques include
\tool{Daikon}~\cite{ErnstDaikon2001} and
\tool{DIDUCE}~\cite{HangalTracking2002}. Such techniques typically require
source code, can struggle to scale, or have other limitations which reduces
their usefulness in complex autonomous systems.
Statistical fault identification techniques use predicates and statistical
methods to localize faults on a source level~\cite{ZhengStatistical2006, ZhengSampled2004, LiblitScalable2005, LiblitIsolation2003}.

General-purpose frameworks for writing and using instrumentation
tools for dynamic binary analysis include \tool{Valgrind} and \tool{Pin}.
Our technique is implemented in \tool{Valgrind} but could generalize to other
frameworks given appropriate engineering effort.
Dynamic binary instrumentation does its work at runtime, allowing it to encompass
any code called by the subject program, whether it be within the original
program, in a library, or elsewhere~\cite{NethercoteValgrind2004, NethercoteValgrind2007, LukPin2005}.
However, these techniques often impose prohibitive runtime overhead; a key
contribution of our approach is a set of engineering mechanisms for enabling their use in
real-time robotics systems.

\paragraph{Intrusion Detection}
Our approach shares conceptual similarities with
techniques for intrusion detection, especially host-based intrusion detection
~\cite{DenningIntrusion1987, WagnerMimicry2002}. 
Intrusion detection models typically monitor 
application interaction with the operating system, particularly in system calls,
seeking abnormal patterns.
Advances on these approaches have included formalizing the system models,
reducing overhead, and incorporating timing as a factor in
patterns~\cite{LuTiming2015}.
To this end, companies have begun applying machine learning and artificial
intelligence techniques in their threat-detection
approaches~\cite{CylanceopticsBrief2018}, as we do in
robotics systems.

\paragraph{Testing Autonomous Vehicles and Robotics}
% SotiropoulosNavigationBugs2017, TuncaliStaliro2016, QuigleyROS2009, TimperleyArdu2018, MadrigalWaymo2017, FraadeBlanarMeasuring2018, KoopmanFramework2018, KoopmanAutonomous2017, TheisslerAutomotive2017, ForrestChallenges2014, HutchisonRobustness2018, RisterModular2007, BeschastnikhDistributed2016
Testing autonomous vehicles and robotics systems presents problems unique
to those domains; and significant challenges exist in ensuring the safety of
autonomous vehicles in an end-to-end fashion~\cite{KoopmanAutonomous2017}. 
Beschastnikh et al. outline the challenges and drawbacks to 
existing approaches to debugging distributed systems, such as robotics
and autonomous vehicles~\cite{BeschastnikhDistributed2016}.
Several approaches have addressed aspects of the problems in testing these
systems.
Sotiropoulos et al. motivate testing robotics in simulation and demonstrate 
the approach's effectiveness in some
domains~\cite{SotiropoulosNavigationBugs2017}.
Tuncali et al. define a \emph{robustness function} for determining how
far a system is from violating its parameters~\cite{TuncaliStaliro2016}.
Notably, however, this approach relies on well-defined system requirements,
which are absent in many systems.
Timperley et al. categorize real bugs reported in the ArduPilot
autonomous vehicle software as to whether they can be reproduced and/or
detected in simulation~\cite{TimperleyCrashing2018}.
Theissler uses anomaly detection on automotive
data~\cite{TheisslerAutomotive2017}, focusing on faults in 
analog vehicle signals rather than program execution, as we do. 
Hutchison et al. outline a framework for robustness testing of robotics and
autonomous systems, highlighting the differences from traditional
software~\cite{HutchisonRobustness2018}.
These unique challenges, and the lack of a single solution for safety, motivate
our proposed technique for detecting errors in robotics systems. 

% CLG: I think these generic "these systems are hard" papers are far enough from
% what we're doing that we don't need to cite them in this paper.  I put some in
% the intro to flesh it out. 
% Koopman and Wagner highlight the significant challenges involved in creating
% an end-to-end process that ensures that autonomous vehicles are designed
% and deployed in such a manner as to take account of all of the myriad concerns
% that contribute to the vehicles' ultimate safety~\cite{KoopmanAutonomous2017}.
% Similarly, Fraade-Blanar et al. establish that there is no uniform definition of
% \emph{safety} as it applies to autonomous vehicles; they propose systematic ways
% to evaluate and improve safety~\cite{FraadeBlanarMeasuring2018}.
% Forrest and Weimer further highlight challenges in detecting and repairing
% faults in certain classes of autonomous systems,
% such as the potential inaccessibility of the
% system, limited computing and power resources, and use of off-the-shelf
% components~\cite{ForrestChallenges2014}.
% \todo{We present X to overcome these challenges.}

A considerable body of work supports formal verification of
cyber-physical systems to avoid faults. 
However, as Zheng et al. point out in their survey of literature
and interviews with practitioners on
verification and validation for cyber-physical systems,
there are many gaps between the
verification work and practical application to entire real
systems~\cite{ZhengVerification2015, ZhengState2017}.  Regardless, these
techniques are orthogonal to the empirical, dynamic techniques we present in
this paper.

\paragraph{The Oracle Problem}
% BarrOracle15, KanewalaOracle2013, PezzeAutomated2014, ShresthaAssertions2011, CoppitSpecification2005, AutomatedStaats2012
One way to view our learned models are as \emph{oracles} of intended program
behavior.
The \emph{oracle problem}~\cite{BarrOracle2015}---determining whether a
program is behaving as intended---is a longstanding problem in software testing
and a significant barrier to automated testing. 
Kanewala and Bieman~\cite{KanewalaOracle2013} survey existing program
testing techniques that attempt to substitute for an oracle.
They highlight several approaches in the domain of computer graphics
that make use of machine
learning~\cite{FrounchiSegmentation2011, ChanMesh2009}.
While one might expect these approaches to be similar to those proposed in
this work, this category of techniques focuses on using machine learning to 
validate program \emph{output}, rather than ensuring its correct
operation at all times.
The distinction is key when applied to situations where properties such
as safety must be maintained at all times.

\section{Conclusions}
\label{sec:discussion}

In this work, we have demonstrated the capability of dynamic binary instrumentation,
when combined with machine learning,
to detect executions that display behavioral errors.
This holds in the relatively noisy context of robotics software
in simulation, an increasingly critical domain. 
As software without closely-defined expected behavior becomes more common, techniques
such as these, that can draw conclusions about whether software is working without needing to know exactly
what it is intended to do, increase in applicability.
Our technique also has the benefit of not being bound by particular language
constraints.  It does not require access to source code, which is often
unavailable in deployed contexts.
Furthermore, the technique requires no semantic
understanding of a program's function to generate a model of the
program's behavior.

Additional work remains in extending this approach to demonstrate its broad
applicability, especially to real, deployed scenarios.  
The approach holds particular promise for detecting errors early, well before
they manifest in obvious crashes.
More engineering work is required to construct tools that are practical to
detect errors early, as well as a more generic approach
that scales to larger systems.

Expanding the types of data collected and simplifying
deployment of techniques like
these in systems in-the-field motivates data sampling to reduce overhead.
This may be supported by a detailed investigation of the trade offs between
sampling frequency and predictive power, and also the number of additional
signals that can be collected without increasing overhead via sampling.
Additionally collecting more signals affords an opportunity to analyze which
signals have the most predictive power and whether those signals are different 
for robotics software as opposed to other types of software.

We have demonstrated the utility of supervised machine learning models for
predicting defective behavior based on low-level signals.  

We have further demonstrated that timing delays can be absorbed into simulated
robotics systems in varying amounts.
These experiments support the observations that overhead
caused by dynamic binary instrumentation does not cause as much runtime
extension as one might expect based on overhead in traditional systems.
In addition, because of inherent nondeterminism in the underlying \acp{SUT},
the changes in behavior caused by delays are often within expected ranges
of behavior under nominal circumstances.
If instrumentation can be calibrated to avoid interfering with critical
points in the software, it is a suitable tool for analyzing \acp{ARS}.

A next step would be to extend the work to make use of
unsupervised learning and novelty-detection algorithms.
These approaches would reduce the need for an oracle for the training data
or the manual annotation burden. 
Such work would involve a deeper investigation of the mathematical properties
of the data and methods that would be likely to distinguish them.

In sum, our techniques based in dynamic binary
instrumentation and machine learning have broad potential applications.
They have the advantages of being language-independent, not requiring
source code, and requiring no semantic understanding of program behavior.
They are effective for finding executions of real-world robotics software that exhibit
errors, an important challenge in our increasingly-automated world.

\section*{Acknowledgment}

This research was partially funded by
AFRL (\#OSR-4066, \#FA8750-16-2-0042) and DARPA (\#FA8750-16-2-0042).
The authors are grateful for their support.
Any opinions, findings, or recommendations expressed are those of the authors
and do not necessarily reflect those of the US Government.

\ifCLASSOPTIONcaptionsoff
  \newpage
\fi



%\bibliographystyle{IEEEtran}
\bibliographystyle{IEEEtran}
\bibliography{IEEEabrv,dsk_ref}

\begin{IEEEbiography}
%[{\includegraphics[width=1in,height=1.25in,clip,keepaspectratio]{figures/deborahkatz}}]{Deborah S. Katz}
Deborah S. Katz recieved the BA degree in Computer Science from Amherst
College, Amherst, Massachusetts, the JD degree from the New York University School of
Law, New York, New York, and the MS and PhD degrees from the Computer Science 
Department at Carnegie Mellon University,
Pittsburgh, Pennsylvania. 
Dr. Katz is interested in techniques for enhancing software quality for autonomous
and robotics systems.
She currently works in research and development at Seegrid.
Dr. Katz completed the work presented here while affiliated with Carnegie
Mellon University.
\end{IEEEbiography}

\begin{IEEEbiography}
%[{\includegraphics[width=1in,height=1.25in,clip,keepaspectratio]{figures/christimperley}}]{Christopher S. Timperley}
Christopher S. Timperley received the MEng and PhD degrees in
Computer Science from the University of York, England. He is a systems scientist
at the Institute for Software Research within the School of Computer Science at
Carnegie Mellon University.
Dr. Timperley is interested in developing methods for building,
enhancing, and assuring software for autonomous and robotics systems.
For more information, please visit \url{http://www.christimperley.co.uk}.
\end{IEEEbiography}

%\newpage

\begin{IEEEbiography}
%[{\includegraphics[width=1in,height=1.25in,clip,keepaspectratio]{figures/clairelegoues}}]{Claire Le Goues}
Claire Le Goues received the BA degree in Computer Science from Harvard University, Cambridge, Massachusetts, and the MS and PhD degrees from the University of Virginia, Charlottesville, Virginia. She is an associate professor with the School of Computer Science, Carnegie Mellon University, where she is primarily affiliated with the Institute for Software Research. She has been recognized by an NSF CAREER Award, the ICSE 2019 Most Influential Paper Award, and the 2020 ACM SIGSOFT Early Career Researcher Award. Dr. Le Goues is interested in constructing high-quality systems in the face of continuous software evolution, with a particular interest in automatic error repair. For more information, please visit \url{http://www.cs.cmu.edu/~clegoues}.
\end{IEEEbiography}


%\vfill


\end{document}